\documentclass[12pt,a4paper]{article}
\usepackage{amsmath,amsfonts,epsfig,color,latexsym}
\usepackage{amssymb}
\usepackage[english, USenglish]{babel}
\usepackage{epsfig,psfrag}
\usepackage{a4wide}
\usepackage{rotating}
\usepackage{slashed}

\usepackage{hyperref}

\csname @addtoreset\endcsname{equation}{section}

\makeatletter
\def\timenow{\@tempcnta\time
  \@tempcntb\@tempcnta
  \divide\@tempcntb60
  \ifnum10>\@tempcntb0\fi\number\@tempcntb
  \multiply\@tempcntb60
  \advance\@tempcnta-\@tempcntb
  :\ifnum10>\@tempcnta0\fi\number\@tempcnta}
\makeatother

\def\oonoo#1#2#3{\vbox{\ialign{##\crcr
	\hfil\hfil\hfil{$#3{#1}$}\hfil\crcr\noalign{\kern1pt\nointerlineskip}
	$#3{#2}$\crcr}}}
\def\oon#1#2{\mathchoice{\oonoo{#1}{#2}{\displaystyle}}
	{\oonoo{#1}{#2}{\textstyle}}{\oonoo{#1}{#2}{\scriptstyle}}
	{\oonoo{#1}{#2}{\scriptscriptstyle}}}

\def\dt#1{\oon{\hbox{\bf .}}{#1}}  
\def\ddt#1{\oon{\hbox{\bf .\kern-1pt.}}#1}
\def\slap#1#2{\setbox0=\hbox{$#1{#2}$}
	#2\kern-\wd0{\hfuzz=1pt\hbox to\wd0{\hfil$#1{/}$\hfil}}}

\begin{document}
\renewcommand{\thefootnote}{\fnsymbol{footnote}}
\newpage
\pagestyle{empty}
\setcounter{page}{0}


\newcommand{\norm}[1]{{\protect\normalsize{#1}}}
\newcommand{\p}[1]{(\ref{#1})}
\newcommand{\half}{{\ts \frac{1}{2}}}
\newcommand \vev [1] {\langle{#1}\rangle}
\newcommand \ket [1] {|{#1}\rangle}
\newcommand \bra [1] {\langle {#1}|}

\newcommand{\cM}{{\cal M}} 
\newcommand{\cR}{{\cal R}} 
\newcommand{\cS}{{\cal S}} 
\newcommand{\cK}{{\cal K}}
\newcommand{\cL}{{\cal L}} 
\newcommand{\cF}{{\cal F}}
\newcommand{\cN}{{\cal N}}
\newcommand{\cA}{{\cal A}}
\newcommand{\cB}{{\cal B}}
\newcommand{\cG}{{\cal G}}
\newcommand{\cO}{{\cal O}}
\newcommand{\cY}{{\cal Y}}
\newcommand{\cX}{{\cal X}}
\newcommand{\cT}{{\cal T}}
\newcommand{\cW}{{\cal W}}
\newcommand{\cP}{{\cal P}}
\newcommand{\nt}{\notag\\} 
\newcommand{\pa}{\partial}
\newcommand{\ep}{\epsilon}
\newcommand{\om}{\omega}
\newcommand{\bep}{\bar\epsilon}
\renewcommand{\a}{\alpha}
\renewcommand{\b}{\beta}
\newcommand{\g}{\gamma}
\newcommand{\s}{\sigma}
\newcommand{\la}{\lambda}
\newcommand{\tl}{\tilde\lambda}
\newcommand{\da}{{\dot\alpha}}
\newcommand{\db}{{\dot\beta}}
\newcommand{\dg}{{\dot\gamma}}
\newcommand{\dd}{{\dot\delta}}
\newcommand{\q}{\theta}
\newcommand{\bq}{\bar\theta}
\newcommand{\bp}{\bar\psi}
\newcommand{\txi}{\tilde\xi}
\newcommand{\bQ}{\bar Q}
\newcommand{\tx}{\tilde{x}}
\newcommand{\tr}{\mbox{tr}}
\newcommand{\+}{{\dt+}}
\renewcommand{\-}{{\dt-}}
\newcommand{\ti}{{\textup{i}}}

\vspace{20mm}

\bigskip
\bigskip

\null\vskip-53pt \hfill
\begin{minipage}[t]{50mm}
CERN-TH-2016-108 \\
LAPTH-021/16
\end{minipage}

\vskip1.1truecm
\begin{center}
\vskip 0.0truecm

 {\Large\bf
 { Composite operators and  form factors in $\cN=4$ SYM }
  }
\vskip 1.0truecm

{\bf  Dmitry Chicherin$^{a}$ and  Emery Sokatchev$^{a,b}$ \\
}

\vskip 0.4truecm
 $^{a}$ {\it LAPTH\,\footnote{Laboratoire d'Annecy-le-Vieux de Physique Th\'{e}orique, UMR 5108},   Universit\'{e} de Savoie, CNRS,
B.P. 110,  F-74941 Annecy-le-Vieux, France\\
  \vskip .2truecm

$^{b}$ Theoretical Physics Department, CERN, CH -1211, Geneva 23, Switzerland 
} \\
\end{center}

\vskip .3truecm

\centerline{\bf Abstract} 
\medskip
\noindent

We construct the most general composite operators of $\cN = 4$ SYM in Lorentz harmonic chiral ($\approx$ twistor) superspace. The operators are built from the SYM supercurvature which is nonpolynomial in the chiral gauge prepotentials. 
We reconstruct the full nonchiral dependence of the supercurvature. We compute all tree-level MHV form factors via the LSZ redcution procedure with on-shell states made of the same supercurvature.

\newpage

\thispagestyle{empty}

{\small \tableofcontents}

\newpage
\setcounter{page}{1}\setcounter{footnote}{0}
\pagestyle{plain}
\renewcommand{\thefootnote}{\arabic{footnote}}

\newpage\setcounter{page}1

\section{Introduction}

In the recent years a lot of effort has been employed for studying scattering amplitudes in the maximally supersymmetric $\cN=4$ gauge theory (SYM). This activity was to a large extent motivated by Witten's approach to amplitudes based on twistor theory \cite{Witten:2003nn}.  New techniques for the computation of amplitudes and  beautiful mathematical structures were discovered. In particular, the superamplitudes have a remarkable Yangian symmetry \cite{Drummond:2008vq,Berkovits:2008ic,Beisert:2008iq,Drummond:2009fd} which strongly suggests that they are integrable. 

Other very interesting objects  in this conformal field theory are the correlation functions of gauge invariant composite operators, in particular of the protected stress-tensor multiplet. Being off-shell quantities, they have a much richer structure than the amplitudes. At the same time, they are finite objects with exact superconformal symmetry. A remarkable connection exists between  scattering amplitudes and the singular light-like  limit of correlators \cite{Alday:2010zy,Eden:2010zz}. This suggests that there is an intimate interplay between the two objects, one on-shell, the other off-shell, and that the conjectured integrable structure of the former somehow extends to the latter. 

There exists a third class of field theory objects, which interpolate between amplitudes and correlators, the form factors of gauge invariant operators.  The form factor in $\cN = 4$ SYM is a quantity which describes the  matrix element of a composite gauge invariant operator $\cO$ (or a supermultiplet of operators) and a final scattering state of particles constituting 
the vector multiplet of $\cN= 4$ supersymmetry,
\begin{align}\label{FFO}
F_{\cal O}(1,2,\ldots,n| x, \q,\bq) = \vev{ 1,2,\ldots,n| \mathcal{O}(x,\q,\bq) |0}\,.
\end{align}
It shares with the amplitude the presence of a number of external on-shell legs. At the same time, like the correlators, it involves an off-shell composite operator. Such a hybrid object is expected to inherit much of the remarkable simplicity of the $\cN=4$ SYM amplitudes, and at the same time to exhibit some of the non-trivial off-shell structure of the correlators. 

Form factors have been extensively studied in a number of papers in the past, for example  at weak coupling \cite{vanNeerven:1985ja,Brandhuber:2010ad,Bork:2010wf,Engelund:2012re} and at strong coupling  \cite{Alday:2007he,Maldacena:2010kp}. We would like to mention in particular Refs.~\cite{Brandhuber:2011tv,Bork:2011cj,Penante:2014sza} where the form factors of the protected half-BPS  operator have been examined and it has been shown that the $\cN=4$ supersymmetry Ward identities determine to a large extent the structure of the MHV form factors. 

Very recently a proposal how to compute the tree-level MHV form factors of all kinds of composite operators in $\cN=4$ SYM was put forward in \cite{Koster:2016loo}. The authors {obtain the form factors from  postulated effective operator vertices. These  are}  non-local objects in twistors space, involving Wilson lines which connect the various constituents of the operator. The local expression is obtained by shrinking the Wilson loop to a point. { Gauge invariance is restored only in this limit.} The main result of the paper is a general formula for the the tree-level MHV form factors.

In the present paper we give  a step-by-step derivation of the tree-level MHV form factors from first principles. We employ the recently proposed formulation of $\cN=4$ SYM in Lorentz harmonic chiral (LHC) superspace \cite{Chicherin:2016fac,Chicherin:2016fbj}. This is an alternative to the twistor space formulation of Mason et al \cite{Mason:2005zm,Boels:2006ir,Adamo:2011cb}. It makes use of the conceptually simpler notion of harmonic superspace, first proposed for the formulation of theories with extended supersymmetry off shell \cite{Galperin:1984av,Galperin:2001uw}. In it one uses harmonic fields  having an infinite expansion on a coset of the R-symmetry group. This expansion provides the infinite sets of auxiliary and pure gauge fields needed  to lift the theory off shell.   The same concept was adapted in \cite{Sokatchev:1995nj} to the Lorentz group instead of the R-symmetry group. It was  used to formulate the  self-dual $\cN=4$ SYM theory of Siegel \cite{Siegel:1992xp} in the form of a Chern-Simons action. In \cite{Chicherin:2016fac,Chicherin:2016fbj} we extended this formulation to the full SYM theory and showed how to compute all non-chiral Born-level correlators of the stress-tensor multiplet, building upon the earlier work in \cite{Chicherin:2014uca}.

In  this  paper we apply the formulation  of \cite{Chicherin:2016fac,Chicherin:2016fbj} to the construction of  composite operators. Some simple examples appeared already in \cite{Chicherin:2016fac}. In \cite{Chicherin:2016soh} we explained the equivalence of our formulation with the alternative  twistor construction in \cite{Adamo:2011dq,Koster:2016ebi}. Here we apply our method to the most general composite operators. We make use of the basic object of the theory, the $\cN=4$ SYM supercurvature $W_{AB}(x,\q,\bq)$. The operators are obtained as local products of supercurvatures and their derivatives.  Only in some  special cases we need to make the definition of the operator non-local in  harmonic space (but not in space-time). In this we differ from the approach of \cite{Koster:2016loo} where all the effective operator vertices are non-local in twistor space.

An important ingredient in the calculation of   form factors is the (super)momentum  on-shell state. In the standard LSZ approach to amplitudes and form factors it corresponds to the amputation of the external legs. Here we carry out this procedure for the supersymmetric propagators that we have found in \cite{Chicherin:2016fac}. We find a very simple and manifestly supersymmetric on-shell state, which we  insert into our definition of the composite operators to obtain the tree-level MHV form factors. Our results agree with those of \cite{Koster:2016loo}.

The paper is organized as follows. In Sect.~2 we review the LSZ procedure for the calculation of form factors and indicate what is needed to supersymmetrize it. Our main point is to use (super)curvatures instead of gauge fields as external states. In Sect.~3 we discuss the $\cN=4$ SYM supercurvature on shell and show how it can be converted into the standard Nair superstate. In Sect.~4 we briefly review the formulation of $\cN=4$ SYM in LHC superspace. In Sect.~5 we explain how to construct composite operators from the supercurvature. The LHC formulation is chiral, so we first consider the chiral truncation of the operators. Then we apply the on-shell $\bQ-$supersymmetry rules found in \cite{Chicherin:2016fac} to reconstruct the full nonchiral operators. Sect.~6 is devoted to the LSZ amputation procedure of the super-propagator, which turns it into an on-shell superstate.  In Sect.~7 we insert the superstate into two simple operators, the stress-tensor and the Konishi multiplets, to obtain explicit examples of form factors. In Sect.~8 we extend our construction to the most general operators with arbitrary spin and twist. This  leads to the most general tree-level MHV form factors. We explain the role of the different gauge frames and the bridges between them in the LHC approach.

\section{Form factors as on-shell limits of correlators}

We study the form factors \p{FFO} using a superspace approach and a supergraph technique. 
In order to specify this quantity we need to construct   the composite operator and the on-shell states of the scattering particles. We are going to express both of them in terms of the $\cN= 4$ {\it nonchiral} supercurvature. 
At this point we slightly deviate from the traditional approach in the amplitude community 
to start with the {\it chiral} on-shell superstate (c.f. \p{css}). Nevertheless, we can obtain the latter by a Grassmann half-Fourier transform, as explained in Sect.~\ref{ncstate}.

Before we embark on the supersymmetric case, let us firstly illustrate our procedure on the simple example of pure YM theory. 
Consider  the operator $\cO=\tr( \tilde F_{\da\db} \tilde F^{\da\db} )$ where $\tilde F$ is the anti-self-dual part of the YM curvature in spinor notation. We wish to evaluate the simplest, tree-level form factor of $\cO$ and a final state with  two positive helicity gluons, 
\begin{align}
\bra{0}g^{(+1)}(p_1)g^{(+1)}(p_2)|\cO(x)\ket{0}\,.
\end{align}
The standard LSZ reduction procedure for calculating the  form factor makes use of the Green's function $\vev{A_{\a\da}(p_1) A_{\b\db}(p_2) \cO(x)}$, in which the two gluon legs are amputated and the gluon states are projected with appropriate polarization vectors onto the required helicities. We prefer to replace the gluons by (self-dual) curvatures. This is commonly used in perturbative QCD calculations. The main advantage is that we maintain gauge invariance at all steps of the calculation. 

So, at tree level we consider the following correlator of YM curvatures (we omit the color indices),
\begin{align} \label{FFcor}
&\vev{ F_{\a\b}(p_1) F_{\gamma\delta}(p_2) \tr(\tilde F_{\da\db} \tilde F^{\da\db})(x) }_{\rm tree} \nt
&
=\int \frac{dq e^{\ti xq}}{(2\pi)^4} \delta^4(p_1+p_2-q)  \vev{F_{\a\b}(p_1) \tilde F_{\da\db}(-p_1) } \vev{F_{\gamma\delta}(p_2) \tilde F^{\da\db}(-p_2) }\nt 
&= \frac{e^{\textup{i} x (p_1 + p_2)}}{p_1^2 p_2^2} (p_1)_{(\a\da} (p_1)_{\b)\db} (p_2)_{(\gamma}^{\da} (p_2)_{\delta)}^{\db}\,.
\end{align}
It is obtained by multiplying together two  free propagators $\vev{F_{\a\b}(p) \tilde F_{\da\db}(-p) }= p_{(\a\da} p_{\b)\db}/p^2$. Notice that the presence of curvatures at the ends of the propagator makes this quantity gauge invariant. 

The next step in the LSZ reduction is to remove the poles by multiplying \p{FFcor} 
by $p_1^2 p_2^2$ and then taking the limit $p^2_i\to 0$. Instead of doing this in the final expression for the correlator \p{FFcor}, we prefer to amputate each propagator separately, i.e. in the middle line of \p{FFcor}. We put the particle momentum on shell, $p_{\a\da} = \la_\a \tl_\da$ and obtain
\begin{align}\label{2.4}
\lim_{p^2\to 0} p^2 \vev{F_{\a\b}(p) \tilde F_{\da\db}(-p) } = \la_\a \la_\b \tl_\da \tl_\db \,.
\end{align}
On shell 
the self-dual curvature of the external state factorizes in a product of negative helicity spinors and 
the creation operator of the gluon,
\begin{align} \label{Ffact}
F_{\a\b}(p) = \lambda_{\a} \lambda_{\b} g^{(+1)}(p)\,.
\end{align}
This allows us to strip off the   helicity spinors 
$\la_\a \la_\b$ from \p{2.4} (this step is equivalent to projecting out  with a polarization vector). In this way we obtain the amputated leg 
\begin{align}\label{ampu}
\vev{g^{(+1)}(p)  \tilde F_{\da\db}(-p) } = \tl_\da \tl_\db \,.
\end{align}
Completing it with the momentum eigenstate wave function $e^{\ti xp}$, we derive the {\it substitution rule}
\begin{align}\label{subst}
 \tilde F_{\da\db}(p_i) \ \Rightarrow \ e^{\ti xp_i}\tl_{i\da} \tl_{i\db}
\end{align}
for each field in the composite operator $\cO=\tr( \tilde F_{\da\db} \tilde F^{\da\db} )$. With this rule we find 
the form factor
\begin{align}\label{FFYM}
 \vev{0| g^{(+1)}(p_1)g^{(+1)}(p_2)| \mathcal{O}(x) |0} = [12]^2 e^{\textup{i}x(p_1 + p_2)}\,,
\end{align}
where  $[12] = (\tl_1)_\da (\tl_2)^\da$.

Let us now come back to the supersymmetric case. The vector multiplet of $\cN=4$ SYM is described by the supercurvature $\mathbb{W}_{AB}(x,\q,\bq)$ (with $A,B= 1 , \ldots,4$). This is a {\it nonchiral} short (half-BPS) superfield.
Its component expansion contains the physical fields: 6 real scalars $\phi^{AB} = \frac{1}{2}\ep^{ABCD} \phi_{CD}$, 4 gluinos $\psi^A_{\a}$ and 4 antigluinos $\bar\psi_{\da A}$, and the two halves of the curvature of the gluon field $( F_{\a\b}, \tilde F_{\da\db})$.
Half-BPS superfields are most naturally described in the R-symmetry harmonic $\cN=4$ superspace. 
One introduces harmonics $w_{\pm}$ for the R-symmetry group $SU(4)$. The projection $\mathbb{W}_{++}$
of the supercurvature  depends on half of the odd variables $\q_+ = w_{+} \cdot \q$, $\bq_+ = \bar w_+ \cdot \bq$ (for more details see Sect.~\ref{ncstate}). 
Then we propose  to consider the correlator 
\begin{align}\label{2.9}
G_n(x,\q,\bq)=\vev{ \mathbb{W}^{\rm free}_{++}(x_1,\q_{1+},\bq_{1+},w_1) 
\ldots \mathbb{W}^{\rm free}_{++}(x_n,\q_{n+},\bq_{n+},w_n)\, \mathcal{O}(x,\q,\bq) } 
\end{align}
as the generalization of the Green's function \p{FFcor}. The role of the YM curvature, which generates the external on-shell states, is now played by the supercurvature $\mathbb{W}^{\rm free}$. It is taken in the free approximation, because the nonlinear (interaction) terms in it do not create single-particle poles as in \p{FFcor}.   
This correlator is gauge invariant because the operator $\cO$ and  the free (linearized) supercurvature 
$\mathbb{W}^{\rm free}_{++}$ are gauge invariant.\footnote{The supercurvature transforms as  $\mathbb{W}_{++} \to e^{- g \Lambda} \mathbb{W}_{++} e^{ g \Lambda}$ under a gauge transformation
with parameter $\Lambda(x,\q_+,\bq_+)$. In the free case $g=0$ it becomes invariant.} The form factor arises as the residue of the
correlator in the on-shell limit, 
\begin{align} \label{C->FF}
F_{\cal O}(1,2,\ldots,n|x,\q,\bq) \sim \lim_{p_1^2, \ldots p_n^2 \to 0} p_1^2 \ldots p_n^2 
\int \prod_{i = 1}^n \frac{d^4 x_i e^{\textup{i} p_i x_i}}{(2\pi)^4}\, G_n(x,\q,\bq)\,.
\end{align}

In Sect.~\ref{ncstate} we explain how the {\it nonchiral} off-shell odd variables $\q_+,\bq_+$ of $\mathbb{W}^{\rm free}_{++}$ 
reduce to the {\it chiral} on-shell Nair's odd variables $\eta_A$,
which serve to assemble the particles of various helicities in a CPT self-conjugate vector multiplet. 
We also show explicitly that the R-symmetry harmonics $w_{\pm}$ in \p{2.9} disappear in the on-shell regime. 
This procedure gives rise to the familiar chiral on-shell superstate (see \p{css}).
We emphasize that the SYM theory is nonchiral, so the chirality of the on-shell superstate 
is not mandatory.\footnote{See \cite{Huang:2011um} for an alternative formulation of the $\cN=4$ SYM amplitudes with nonchiral superstates.} Further, in Sect.~\ref{state} we carry out the amputation of the superspace propagator involving $\mathbb{W}^{\rm free}_{++}$ to obtain the analog of the on-shell state \p{ampu} and of the substitution rule \p{subst}. Finally, in Sects.~\ref{MHV}--\ref{compFF} we apply these rules to the computation of various form factors.

\section{From the nonchiral supercurvature to the chiral on-shell superstate} \label{ncstate}

The supercurvature $\mathbb{W}_{AB}(x,\q,\bq)$ is restricted by a number of  constraints 
that put its component fields on shell \cite{Sohnius:1978wk}.
The constraints can be (partially) solved in an $SU(4)$ covariant manner in RH superspace\footnote{In this paper we employ two abbreviations, LH for Lorentz harmonics and RH for R-symmetry harmonics. The former parametrize a coset  of the chiral half of the Euclidean Lorentz group $SU(2)_{L} \times SU(2)_{R}$, the latter describe a coset of the R-symmetry group $SU(4)$.} \cite{Hartwell:1994rp,Heslop:2000af} (for a recent review see \cite{Eden:2011yp}).
We introduce a set of harmonics $w^{a}_{+ A}, \, w^{a'}_{- A}$ and their conjugates 
$\bar w^A_{-a}, \, \bar w^A_{+a'}$ on the R-symmetry group $SU(4)$, 
projecting the index $A=1,\ldots,4$ of the (anti)fundamental irrep onto the subgroup $SU(2)\times SU(2)'\times U(1)$ (indices $a,a',\pm$)\footnote{We raise and lower  the R-symmetry indices $a,a'$, as well as the Lorentz indices $\a,\da$ with the help of the two-dimensional
Levi-Civita tensors $\ep_{ab}$, $\ep^{a'b'}$, $\ep_{\a\b}$, $\ep^{\da\db}$, etc. Our convention is $\ep_{ab}\ep^{bc} = \delta_a^c$, $\ep_{12}=1$.}: 
\begin{align}
&(\ w^a_{+A} \ ,\ w^{a'}_{-A}\ ) \in SU(4): \notag \\ \label{5.1}
&w^{a}_{+ A}   \bar w^{A}_{- b} = \delta^{a}_{b}  \,, \quad w^{a}_{+ A}  \bar w^A_{+a'} = 0\,, \quad w^{a'}_{- A}  \bar w^{A}_{- b} =0\,, \quad w^{a'}_{- A}  \bar w^A_{+b'} = \delta^{a'}_{b'}\nt
&w^{a}_{+ A} \bar w^B_{-a} +  w^{a'}_{- A} \bar w^B_{+a'} =\delta^B_A \, .
\end{align}
The functions of the RHs are 
covariant with respect to the coset subgroup. In particular, this implies homogeneity in the $U(1)$ charge. 
We use RHs to project the $SU(4)$ indices carried by the odd variables and the fields. 
For example,
\begin{align}
\phi_{++} = \frac{1}{2}\ep^{a'b'} \bar w^{A}_{+a'} \bar w^{B}_{+b'} \phi_{AB} \;\;,\;\;
 \psi^a_{\a +} =  w^a_{+A}\psi^{A}_{\a} \;\;,\;\; \bar\psi_{\da a'+} = \bar w^A_{+a'} \bar\psi_{\da A}\,.
\end{align}

The constraints imposed on the supercurvature $W_{AB}$ can be partially solved in the linearized (or free) approximation. 
We introduce its RH projection onto the highest weight state of the irrep $[010]$ of $SU(4)$, 
$\mathbb{W}_{++} = \frac{1}{2}\ep^{a'b'}  \bar w^{A}_{+a'} \bar w^{B}_{+b'} \mathbb{W}_{AB}$. Then we  interpret part of the constraints as an R-analyticity condition, which is an example of Grassmann analyticity \cite{Galperin:1984av}. 
This means that $\mathbb{W}^{\rm free}_{++}$
depends only on the $\q_+,\bq_+$ projections of the odd variables,
\begin{align}
 \q_{+\a}^a = w^{a}_{+ A} \q^A_\a\,, \qquad  
\bq^\da_{+a'} = \bar w^A_{+a'}\bq^\da_{A}\,,
\end{align}
but does not depend on their conjugates $\q_-,\bq_-$.  
In the free approximation the supercurvature is an ultrashort superfield \cite{Heslop:2000af,Ferrara:2000eb}, i.e. 
the expansion of $\mathbb{W}^{\rm free}_{++}$ in the odd variables contains only the terms $(\q_+)^k (\bq_+)^m$ with $k,m \leq 2$ 
\cite{Andrianopoli:1999vr},
\begin{align}\label{Wexp}
\mathbb{W}^{\rm free}_{++}(x,\q_+,\bq_+,w) &= \phi_{++} + \q^{\a a}_+ \psi_{\a a +} +  (\q_+^\a\cdot \q_+^\b)F_{\a\b} \nt 
&+ \bq^{\da a'}_{+} \bar\psi_{\da a'+} + \ldots + \bq^{\da a'}_+(\q^{\a}_{+} \cdot \q^{\b}_{+})  \pa_{\a\da}\psi_{\b a'-} \nt
&+ (\bq^{\da}_{+} \cdot \bq^{\db}_{+}) \tilde F_{\da \db} + \ldots +  
(\q^{\a}_{+} \cdot \q^{\b}_{+}) (\bq^{\da}_{+} \cdot \bq^{\db}_{+}) \pa_{\a\da} \pa_{\b\db} \phi_{--}\ .
\end{align} 
Each term in the expansion carries $U(1)$ charge $(+2)$.\footnote{We use spinor units of charge, hence the charge $(+2)$ of the HWS of the vector irrep of $SO(6)\sim SU(4)$.}
The dots stand for other terms with derivatives of scalars and (anti)-gluinos.
The physical fields carrying $SU(4)$ indices are split up into a number of RH projections,
\begin{align}
\phi_{++} \;,\; \phi_{--} \;,\; \phi^{aa'}_{+-} \;,\; \psi_{\a a+} \;,\; \psi_{\a a'-}
\;,\; \bar\psi_{\a a'+} \;,\; \bar\psi_{\a a-}
\end{align}
which appear as various components of the ultrashort nonchiral multiplet $\mathbb{W}^{\rm free}_{++}$ \p{Wexp}.

The remaining  constraint defines   $\mathbb{W}_{++}$ as a highest-weight state of $SU(4)$  (harmonic analyticity). It  restricts the RH dependence to polynomial and at the same time puts the component fields from \p{Wexp} on shell. 

On the mass shell we can pass from fields to momentum eigenstates. The gauge-covariant fields reduce to products of an on-shell state  and helicity 
spinors carrying the Lorentz indices (cf. \p{Ffact}), 
\begin{align} \label{states}
\quad F_{\a\b} = \la_\a \la_\b\, g^{(+1)} \;\;,\;\; \psi^{A}_{\a}= \la_\a\, \psi^{(+\frac{1}{2})A}
\;\;,\;\; \bar\psi_{\da A}= \tilde\la_\da\, \psi^{(-\frac{1}{2})}_{A} 
\;\;,\;\; \tilde{F}_{\da\db} = \tilde\la_{\da} \tilde\la_{\db}\, g^{(-1)}\,.
\end{align} 
Here the creation operators $g^{(+1)}$, etc. are supposed to act on the vacuum.
After replacing the fields in \p{Wexp} by the on-shell states \p{states}, we see that the odd variables get projected by the 
helicity spinors,
\begin{align} \label{3.7}
\chi^a_+ \equiv \q^{\a a}_+\la_\a \qquad, \qquad \eta^{a'}_+ \equiv \bq^{\da a'}_+ \tilde\la_\da\,,
\end{align} 
so the supercurvature $\mathbb{W}^{\rm free}_{++}$ is converted into the superstate 
\begin{align} \label{Wonshell}
\Phi_{++}(p,\chi_+ ,\eta_+,w) = &
\phi_{++} + \chi^{a}_+ \psi^{(+\frac{1}{2})}_{a +} +  (\chi_+)^2 g^{(+1)} 
+ \eta^{a'}_{+} \psi^{(-\frac{1}{2})}_{a'+} + (\eta_{+})^2 g_{(-1)} \nt 
&+ (\chi_+)^2 \eta^{a'}_+ \psi^{(+\frac12)}_{a'-} + \ldots + (\chi_+)^2 (\eta_+)^2 \phi_{--}
\end{align} 
where we have omitted the same terms as in \p{Wexp}. This is a nonchiral realization of the on-shell state
which uses RHs to maintain the manifest $SU(4)$ invariance (see \cite{Huang:2011um}). 

In order to rewrite the on-shell state \p{Wonshell} in the more familiar chiral 
form, we perform a Fourier transform (FT)  from $\chi^a_{+}$ to $\eta_{-a}$. The resulting change of variables 
$(\chi_{a+},\eta_{a'+}) \to (\eta_{a-},\eta_{a'+}) \equiv \eta_A$ restores the $SU(4)$ index,
\begin{align} \label{FT}
\Phi(p,\eta) \equiv \Phi(p,\eta_-,\eta_+) = \int d^2 \chi_+ e^{\eta_-\chi_+ } \Phi_{++}(p,\chi_+ ,\eta_+,w)\,.
\end{align} 
As an illustration how the RHs drops out after the FT \p{FT}, consider
the terms in   \p{Wonshell}  containing the projections of $\psi^{(+\frac12)}$. They are transformed into 
$\eta^{a}_- \psi^{(+\frac{1}{2})}_{a +} + \eta^{a'}_+ \psi^{(+\frac12)}_{a'-} = \eta_A \psi^{(+\frac12)A}$ 
in view of the completeness relation for RHs (the third line in \p{5.1}).
In the same way, in all the terms in \p{Wonshell} 
we reconstruct the representations of $SU(4)$ from 
their projections and obtain the familiar chiral on-shell superstate \cite{Nair:1988bq} 
\begin{align}\label{css}
\Phi(p,\eta) = g^{(+1)} +\eta_A \psi^{(+\frac12)A} +\frac{1}{2}\eta_A \eta_B \phi^{AB}+
(\eta^3)^{D} \psi^{(-\frac12)}_{D} + (\eta)^4 g^{(-1)}\,,
\end{align} 
where 
\begin{align} \label{etapow}
(\eta^3)^{D} = \frac{1}{3!} \ep^{ABCD} \eta_{A}\eta_{B}\eta_{C} \;\;,\;\;
(\eta)^4 = \eta_1\eta_2\eta_3\eta_4\,.
\end{align}

\section{$\cN = 4$ SYM in Lorentz harmonic chiral superspace} \label{LHC}

According to \p{C->FF}, the form factors are obtained as the on-shell residues  
of the correlator of the supercurvatures. In \cite{Chicherin:2016fac} we proposed a construction of the supercurvatures and an off-shell formulation of $\cN = 4$ SYM in  Lorentz harmonic chiral (LHC) superspace. 
In this formulation the chiral half of   $\cN = 4$ supersymmetry is realized off shell. We have developed a Feynman supergraph technique with manifest  chiral supersymmetry.
Having half of the supersymmetry off shell is possible due to 
the infinite number of auxiliary and pure gauge fields of arbitrarily high spin.
Here we briefly review this formalism.

We work with the Euclidean Lorentz group $SO(4) \sim SU(2)_{L} \times SU(2)_{R}$. The left and right factors act on the undotted and dotted Lorentz indices of the space-time coordinates $x^{\da \a}= x^\mu\tilde\sigma^{\da\a}_\mu $, respectively.
The LH variables $u^{+}_{\a}$ and $u^-_{\a}$ are a pair of spinors forming an $SU(2)_L$ matrix \cite{Galperin:2001uw,Devchand:1992st}:
\begin{equation}\label{2}
\left(u^+_{\a} , u^-_{\a} \right) \in SU(2)_L\ : \ u^{+\a}u^-_{\a} = 1 \ , \ (u^+_{\a})^* = - u^{-\a} \ , \ (u^{+\a})^* = u^{-}_{\a}\,.
\end{equation}
The LHs project the fundamental representation of $SU(2)_L$ onto the $U(1)$ subgroup, so that 
their indices $\pm$ denote the $U(1)$ charge. All expressions have to be homogeneous in the $U(1)$ charge.
To distinguish the charges $\pm$ of the LHs from those of the RHs from Sect.~\ref{ncstate}, we indicate the former upstairs and the latter downstairs.

Our (super)fields  are LH functions defined by their infinite LH expansion on $S^2 \sim SU(2)_L/U(1)$. In it we find irreducible representations of $SU(2)_L$ 
of arbitrarily high spin (totally symmetric multispinors). For example, for a charge $(+2)$ LH field   we have
\begin{equation}\label{3}
f^{++}(x,u) = f^{\a\b}(x) u^+_{\a} u^+_{\b} + f^{(\a\b\gamma)}(x) u^+_{\a} u^+_{\b} u^-_{\gamma} 
+ f^{(\a\b\gamma\delta)}(x) u^+_{\a} u^+_{\b} u^-_{\gamma} u^-_{\delta} + \ldots
\end{equation}
So, an LH field consists of an 
infinite set of ordinary multispinor fields $f^{(\a_1 \ldots \a_{m})}(x)$.

The differential operators compatible with the normalization condition $u^{+\a}u^-_{\a} = 1$ \p{2} 
are the LH derivatives
\begin{align}\label{4}
\pa^{++}= u^{+\a} \pa / \pa u^{-\a} \ , \ \ \
\pa^{--}= u^{-\a} \pa / \pa u^{+\a}\,.
\end{align}
Acting on an LH function  they increase and decrease, respectively, its $U(1)$ charge. 
Supplementing them with the Cartan charge $\pa^0$ which counts the $U(1)$ charge of the LH functions
we obtain the algebra of $SU(2)_L$,
\begin{align}\label{su2}
[\pa^{++}, \pa^{--}] = \pa^0 \;,\qquad [\pa^0, \pa^{++}] = 2 \pa^{++} \;,\qquad [\pa^0, \pa^{--}] = -2 \pa^{--} \,.
\end{align} 
The restriction to functions on $SU(2)_L$ with definite charge gives a particular realization of the LH coset $SU(2)_L/U(1)$. 

An LH function with $U(1)$ charge $q \geq 0$, which is the highest weight of a finite-dimensional irrep of $SU(2)_L$ of spin $q/2$, 
is a multispinor of rank $q$. This property can be  equivalently formulated as an LH differential equation,  
\begin{align}\label{5}
q\geq 0: \ \  \pa^{++}f^{(q)}(u) = 0 \ \ \Rightarrow \ \ f^{(q)}(u) = f^{\a_1 \ldots \a_{q}} u^+_{\a_1} \ldots u^+_{\a_{q}}\,. 
\end{align}

We also need an $SU(2)_L$ invariant LH integral on $S^2$. 
For the LH functions of nonzero $U(1)$ charge the integration gives zero, 
and for the chargeless LH functions having the LH expansion 
$f(u)=f + f^{(\a\b)} u^+_{\a} u^-_{\b}+\ldots$ the integral 
picks the singlet part $\int du\; f(u) = f$.
In particular, $\int d u = 1$. 
This rule is compatible with integration by parts for the LH
derivatives \p{4}. Alongside with the regular LH functions 
that admit LH expansions on $S^2$ (see \p{3}), we also consider singular LH distributions.
The LH delta function $\delta(u,v)$ is defined by the property
\begin{align}\label{delta}
\int dv\; \delta(u,v)\; f^{(q)}(v) = f^{(q)}(u)\, 
\end{align}
with a test function of $U(1)$ charge $q$.

The LHs are used to project the Lorentz indices of the odd variables and derivatives,
\begin{align}\label{dif}
\q^{\pm A} = u^{\pm}_{\a}\q^{\a A} \ \ \ , \ \ \
\pa^\pm_A \equiv u^{\pm \a} \pa / \pa \q^{\a A}\ \ \ , \ \ \ 
\pa^{\pm}_\da \equiv u^{\pm \a} \pa/\pa x^{\da\a}\,.
\end{align}

In this section we keep only the {\it chiral} odd variables $\q^{A}_{\a}$ and work with 
superfields which transform covariantly with respect to the $Q$-half of  the $\cN = 4$ 
supersymmetry algebra. The odd variables $\bq^{\da}_{A}$ are absent, so 
the $\bQ$-half of  supersymmetry is not manifest.
We extensively use {\it L-analytic} harmonic superfields $\Phi(x,\q^+,u)$
which depend only on half of the chiral odd variables. Equivalently we can
formulate the L-analyticity of an LH superfield as $\pa^+_{A} \Phi = 0$ (see \p{dif}). 
L-analyticity is another form of  Grassmann analyticity (cf. the R-analyticity from Sect.~\ref{ncstate}).

In the gauge theory we consider gauge transformations whose parameter is an  
L-analytic harmonic superfield of $U(1)$ charge zero $\Lambda(x,\q^+,u)$ in the adjoin representation of the gauge group $SU(N_c)$. 
This is the so-called {\it analytic gauge frame}.
In it the flat derivatives $\pa^{++}$, $\pa^{--}$, $\pa^-_{A}$, $\pa^{\pm}_{\da}$ (see \p{4}, \p{dif}) 
are extended to  
covariant derivatives $\nabla^{++}$, $\nabla^{--}$, $\nabla^-_{A}$, $\nabla^{\pm}_{\da}$ by adding  gauge connections.\footnote{In our conventions the gauge connections are accompanied by the YM coupling $g$,
$\nabla \equiv \pa + g A$. The infinitesimal gauge transformations have the form $\delta_{\Lambda} A = \nabla \Lambda$ with an L-analytic parameter $\Lambda$.} 
The derivatives $\pa^0$ and $\pa^+_{A}$ remain flat since $\pa^0 \Lambda = \pa^+_{A} \Lambda =0$.
A key role is played by the gauge connections $A^{++}(x,\q^+,u)$ and $A^{+}_{\da}(x,\q^+,u)$, 
which are L-analytic superfields. In \cite{Chicherin:2016fac} we identified them as the dynamical fields of $\cN = 4$ SYM.
They are {\it gauge prepotentials}, i.e. the remaining gauge connections $A^{--}$, $A^-_A$, $A^-_{\da}$  
as well as the supercurvatures
can be expressed in their terms. The latter are not L-analytic, 
so they depend on the full chiral odd variable $\q^{A}_{\a}$.
 
The crucial step in constructing all the gauge connections in terms of the dynamical fields $A^{++}$ and $A^+_{\da}$ 
is finding $A^{--}$. The $SU(2)_L$ algebraic structure of the LH derivatives provides the key.
Indeed, the gauge connection $A^{--}$ is present in the covariantized commutation relation \p{su2},
\begin{align}\label{covarr}
[\nabla^{++}, \nabla^{--}] = \pa^0\ .
\end{align}
This is an LH differential equation on $S^2$ with a unique solution for $A^{--}$.
The solution is a chiral superfield (not L-analytic) which is 
nonpolynomial in $A^{++}$ \cite{Zupnik:1987vm,Galperin:2001uw},
\begin{align}\label{448}
A^{--}(x,\q,u) =  -\sum^\infty_{n=1} (-g)^{n-1}\int du_1\ldots du_n\; { 
A^{++}(x,\q \cdot u^+_1,u_1) \ldots A^{++}(x,\q \cdot u^+_n,u_n) \over (u^+u^+_1)(u^+_1u^+_2) \ldots
(u^+_nu^+)}
\end{align} 
with $\q^{A} \cdot u^+_k = \q^{\a A} (u^+_{k})_{\a}$.
The right-hand side of \p{448} is local in $(x,\q)$ space but nonlocal in LH space.
It has zero $U(1)$ charge with respect to the integration LH variables $u_1, \ldots, u_n$ and charge $(-2)$ with respect to $u$.

In the LH formulation the action of the $\cN=4$ SYM involves the dynamical fields $A^{++}$ and $A^+_{\da}$. 
It consists of two terms, $S_{\cN=4} = S_{\rm CS}+S_{\rm Z}$. The first term is a Chern-Simons-like action which
describes the self-dual sector of the theory \cite{Siegel:1992xp,Sokatchev:1995nj,Witten:2003nn}. 
The second term involves only $A^{++}$ and it is nonlocal in LH space (it looks very similar to \p{448}, see \p{Zup} below).  The role of this term is to complete the self-dual sector to the full SYM theory \cite{Mason:2005zm}. 
The form of $S_{\rm Z}$ coincides with the $\cN=2$ SYM action in RH superspace as given by Zupnik \cite{Zupnik:1987vm}.

In the quantum theory we need a gauge fixing condition. 
Following \cite{Boels:2006ir,Cachazo:2004kj} we choose the light-cone (or `axial' or `CSW') gauge $\xi^{\+ \da} A^+_{\da} = 0$. It is defined by the auxiliary 
LHs $\xi^{\dt\pm}$ for the $SU(2)_R$ factor of the Euclidean Lorentz group (called `reference spinor' in  \cite{Boels:2006ir}). 
To distinguish them from the LHs $u^\pm$ on $SU(2)_L$ (see \p{2}) we denote their $U(1)$ charge by $\dt\pm$.
In this gauge the $S_{\rm CS}$ part of the action becomes free and all interactions are due to $S_{\rm Z}$.
For the calculation of form factors we need propagators in the momentum representation.
Repeating the argument in \cite{Chicherin:2016fac} (see also \cite{Jiang:2008xw}), but this time in momentum space, we obtain
\begin{align}
&\langle A^{+}_{\da}(p,\q^+,u_1) A^{+\db}(-p,0,u_2) \rangle=0\,, \label{prop3} \\
&\vev{A^{++}(p,\q^+,u_1) A^{++}(-p,0,u_2)}= 4\pi \delta^2( p^{\- +} )\, \delta (u_1, u_2)\, \delta^{4}(\q^+)\,, \label{prop1}\\
&\langle A^{+}_{\da}(p,\q^+,u_1) A^{++}(-p,0,u_2) \rangle=
2\ti \xi^{\-}_{\da} / p^{\- +}  \,
\delta(u_1,u_2) \,\delta^{4}(\q^+)\,.  \label{prop2} 
\end{align}
The LH delta function is defined in \p{delta}. The fermionic delta function
$\delta^{4}(\q^+) \equiv \q^+_{1} \q^+_{2} \q^+_{3} \q^+_{4}$ carries $U(1)$ charge $(+4)$.
The right-hand side of \p{prop1} contains also the complex delta function
$\delta^2(t) \equiv \delta(t,\bar t)\,$ satisfying the relation 
${\partial\over \partial\bar t}\;\frac1{t} = \pi \delta^2(t)$. 
More specifically, we use $ \delta^2( p^{\- +} ) =  \delta( p^{\- +},p^{\+ -} )$
where $p^{\dt\pm \mp} \equiv \xi^{\dt\pm}_{\da} p^{\da \a} u^{\mp}_{\a}$. 

\section{Operator supermultiplets in LHC superspace} \label{opers}

We construct composite operators from the supercurvature $\mathbb{W}_{AB}(x,\q,\bq)$.
It appears in the anticommutator of the covariantized spinor derivatives
$\{ \nabla^{\a}_{A} , \nabla^{\b}_{B} \} = \ep^{\a\b} g \mathbb{W}_{AB}$. \footnote{The vanishing of the symmetric in $\a\b$ part of  the right-hand side  defines $\cN=4$ SYM.}
In this Section we consider several supermultiplets of composite operators 
which are of prime interest for the applications.
These are the Konishi multiplet (and its higher twist generalizations) 
and the half-BPS multiplets (including the stress-tensor multiplet).
We explain how to construct them in terms of the L-analytic superfields (see Sect.~\ref{LHC}).
The construction is particularly simple for the chiral truncation of the multiplets.
This is not a surprise since the constituent L-analytic superfields are chiral. 
The nonchiral sector of the multiplets is much more involved.
We construct it acting on the chiral sector with the $\bar Q$-supersymmetry generators.
In Sects.~\ref{state} and \ref{MHV} we show that the complications which arise in the nonchiral sector 
are removed if we restrict ourselves to the tree level MHV form factors.
These simplest form factors do not capture all the sophisticated details of the nonchiral composite operators.
In Sect.~\ref{allop} we generalize our construction to composite operators of arbitrary spin and twist.

\subsection{Chiral truncation of the multiplets}

Let us start with the chiral truncation of the supercurvature $W_{AB}(x,\q) \equiv \mathbb{W}_{AB}|_{\bq = 0}$. 
In the analytic frame (see Sect.~\ref{LHC}) it is an LH superfield of $U(1)$ charge zero, so
the defining anticommutation relation takes the form 
$\{ \nabla^{-}_{A} , \pa^{+}_{B} \} = g W_{AB}(x,\q,u)$ where $\nabla^{-}_{A} = \pa^-_A + g A^-_A(x,\q,u)$.
In \cite{Chicherin:2016fac} we expressed the gauge connection $A^-_A$ in terms of the gauge prepotential $A^{++}$ and found 
a concise form of the chiral supercurvature (see \p{448}),
\begin{align}\label{Wab}
W_{AB}(x,\q,u) = \pa^+_{A} \pa^+_{B} A^{--}\,.
\end{align}
It is covariant with respect to gauge transformations with an L-analytic parameter $\Lambda$, 
and it is covariantly LH independent,  
\begin{align}\label{314}
W_{AB} \to  e^{-g\Lambda(x,\q^+,u)} W_{AB}\, e^{g\Lambda(x,\q^+,u)}\ \ \ , \ \ \ 
\nabla^{++} W_{AB}(x,\q,u)  = 0\,.
\end{align}

We can construct multiplets of operators multiplying together several chiral supercurvatures $W_{AB}$ and taking 
the trace over the adjoint representation of the gauge group $SU(N_c)$.\footnote{In this paper we consider only single-trace operators. The generalization to multi-trace operators is straightforward.}
For example, the chiral truncation of the Konishi multiplet is
\begin{align} \label{konishi}
K(x,\q,u) = \ep^{ABCD} \tr \left( W_{AB} W_{CD} \right)\,.
\end{align}
The supercurvatures are taken in the analytic frame, i.e. they depend on the LHs.
However, the gauge invariant operators like \p{konishi} are LH independent.
Indeed, due to \p{314} we have $\pa^{++} K(x,\q,u) = 2 \ep^{ABCD} \tr \left( W_{AB} \nabla^{++} W_{CD}\right) = 0$.
Then in view of lemma \p{5} we have $K(x,\q,u) = K(x,\q)$. 

Evidently, the $SU(4)$ singlet structure in \p{konishi} is not the only possibility. 
Another interesting subclass of operators are the chiral truncated half-BPS multiplets. Their bottom components transform in the irrep of the R-symmetry group with Dynkin labels $[0,k,0]$,
\begin{align} \label{halfBPS}
\cO_{[0,k,0]}(x,\q_+,w) = \tr \left( W^k_{++}\right) \ \ \ , \ \ \ W_{++} = (\pa^+_+)^2 A^{--}  \,.
\end{align}
Here $ W_{++}(x,\q,u,w)$ is the RH projection of the supercurvature \p{Wab} on the
highest weight state 
$W_{++} =  \frac{1}{2}\ep^{a'b'} \bar w^{A}_{+a'} \bar w^{B}_{+b'} W_{AB}$, and $\pa^+_{+a'} \equiv u^{+\a}\pa_{\a A} \bar w^{A}_{+a'}$.
In this example we have to deal with RHs of $SU(4)$ and LHs of $SU(2)_L$ simultaneously.
The LH independence of the gauge invariant operator \p{halfBPS} is established 
in the same way as for the Konishi multiplet. Let us note that $W_{++}$, contrary to the gauge invariant half-BPS operators 
\p{halfBPS}, is not R-analytic (see Sect.~\ref{ncstate}), i.e. $W_{++}$ depends on both $\q_+$ and $\q_-$.
In fact the dependence on $\q_-$ takes the form of a generalized gauge transformation\footnote{The reason
is that in the analytic frame the $\q_-$-independence condition involves covariant derivatives, 
$\pa^+_+ W_{++} = \nabla^-_+ W_{++} = 0$. This can be changed by going to another, R-analytic frame where $\nabla^\a_+=\pa^\a_+$ but there $\nabla^+_-$ becomes covariant.}, so it drops out in 
the gauge covariant combination \p{halfBPS}. Also $\q_-$ drops out from $W^{\rm free}_{++}$ 
in the free approximation (recall \p{Wexp}), which is insensitive to the choice of gauge frame.

The chiral truncated supercurvature contains only the scalars $\phi_{AB}$, the gluinos $\psi^{A}_{\a}$ 
and the self-dual YM curvature $F_{\a\b}$ (the first line in \p{Wexp}). So only this subset of fields appears in the 
multiplets \p{konishi} and \p{halfBPS}. The anti-gluinos $\bar\psi^{\da}_A$ and the anti-self-dual YM curvature $\tilde F_{\da\db}$ reside in the 
nonchiral sector of the supercurvature (the last two lines in \p{Wexp}). 

\subsection{Complete nonchiral multiplets}\label{s5.2}

Since the odd $\bq$ variable is absent in the LHC formulation of $\cN = 4$ SYM, 
we have to use $\bQ$-supersymmetry to restore the nonchiral sector of the supercurvature \p{Wab},
\begin{align}\label{WbQ}
\mathbb{W}_{AB}(x,\q,\bq,u) = e^{\ti\bq \cdot \bQ}\, W_{AB}(x,\q,u) \,, \qquad \bq \cdot \bQ \equiv \bq^{\da}_{A}  \bQ^{A}_{\da} \,.
\end{align}
Unlike $Q$-supersymmetry, $\bQ$-supersymmetry is not manifest. It is realized on the dynamical fields
$A^{++}$ and $A^+_{\da}$ in the following way \cite{Chicherin:2016fac} (our generators act on the fields, not on the coordinates),
\begin{align}
&\bQ^B_\db A^+_\da  =  -2 \q^{+B} \pa^-_\db A^+_\da + (\bQ_{\rm Z})^B_\db A^+_\da \ \ \ ,\ \ \  
\bQ^B_\db A^{++} = -2 \q^{+B} \bigl( \pa^-_\db A^{++} + A^+_\db \bigr) \,. \label{bQ}
\end{align} 
The $\bQ$-variations mix up both dynamical fields $A^{++}$ and $A^+_{\da}$. 
They act highly nontrivially on $A^+_{\da}$ due to the term $\bQ_{\rm Z}$ which we do not write out explicitly here. It involves the gauge connection $A^{--}$ defined in \p{448}, so it is  nonpolynomial in $A^{++}$.
The supersymmetry algebra closes only on shell and modulo gauge transformations.
The $\bQ_{\rm Z}$-term is irrelevant for MHV tree level form factors (see the Appendix). 
In what follows we shall drop $\bQ_{\rm Z}$. The remaining part of $\bQ$-supersymmetry \p{bQ} corresponds to the 
self-dual sector of $\cN = 4$ SYM \cite{Siegel:1992xp,Sokatchev:1995nj,Witten:2003nn}. In this simplified case we can define the on-shell nonchiral superfield
\begin{align}\label{5.7}
\mathbb{A}^{++}(x,\q^+,\bq,u) = e^{\ti \bq \cdot \bQ}\, A^{++}(x,\q^+,u)
\end{align}
where we repeatedly apply the $\bQ$-variations \p{bQ},
\begin{align}\label{multbQ}
\bQ^{A_1}_{\da_1} \ldots \bQ^{A_k}_{\da_k} A^{++} 
= (-2)^k \q^{+A_1} \ldots \q^{+A_k} \bigl( \pa^-_{\da_1} \ldots \pa^-_{\da_k} A^{++} + k \pa^-_{(\da_1} \ldots \pa^-_{\da_{k-1}} A^+_{\da_k)} \bigr)  
\end{align}
for $k=1,\ldots,4$. The fifth variation vanishes, since $(\q^+)^5 = 0$. 
Using the notion of $\mathbb{A}^{++}$ we recast the supercurvature \p{WbQ} in the following form (recall \p{Wab})
\begin{align}\label{ncW}
&\mathbb{W}_{AB}(x,\q,\bq,u) = \pa^+_A \pa^+_B \mathbb{A}^{--} \;,
\end{align}
{defining the nonchiral analog of $A^{--}$ \p{448}}
\begin{align} 
&\mathbb{A}^{--}(x,\q,\bq,u) = e^{\ti \bq \cdot \bQ}\, A^{--}(x,\q,u) \,, \nt 
&\mathbb{A}^{--}(x,\q,\bq,u)  = -\sum^\infty_{n=1} (-g)^{n-1}\int du_1\ldots du_n\; { 
\mathbb{A}^{++}(1) \ldots \mathbb{A}^{++}(n) \over (u^+u^+_1)(u^+_1u^+_2) \ldots
(u^+_nu^+)} \label{ncA--}
\end{align}
with $\mathbb{A}^{++}(k) \equiv \mathbb{A}^{++}(x,\q \cdot u^+_k, \bq ,u_k)$.

The nonchiral completion of the half-BPS operators \p{halfBPS} is obtained from the 
RH projection of the nonchiral supercurvature \p{ncW},
\begin{align} \label{W++}
\mathbb{W}_{++}(x,\q,\bq,u,w) = 1/2\, \ep^{a'b'}  \bar w^{A}_{+a'} \bar w^{B}_{+b'} \mathbb{W}_{AB}(x,\q,\bq,u)\,.
\end{align}
Just like its chiral counterpart in \p{halfBPS}, $\mathbb{W}_{++}$ in \p{W++} is not R-analytic (see Sect.~\ref{ncstate}), i.e.
it depends not only on $\q_+,\bq_+$ but also on $\q_-$, $\bq_-$. 
The reason is that we work in the analytic frame and we have constructed $\mathbb{W}_{++}$ from the L-analytic fields $A^{++}$, $A^+_{\da}$. In \p{W++} 
$\q_-$ and $\bq_-$ appear in the form of a generalized gauge transformation, so they drop out in  gauge-invariant quantities.

  \section{The supercurvature $\mathbb{W}_{++}$ as an on-shell state}\label{state}
  
Let us briefly recall  the construction of the tree-level form factor in  Sect.~2. We used the amputated propagator \p{2.4} between the self-dual and anti-self-dual YM curvatures. It gave us the external  on-shell state \p{ampu}, which we substituted for each field $\tilde F$ in the composite operator $\cO=\tr(\tilde F \tilde F)$. The result was the tree-level form factor \p{FFYM}. 
 
 In this section we  repeat the argument in the supersymmetric case.  We start by replacing the curvature $F$ at the external legs by the  free supercurvature \p{ncW}, \p{W++}
  \begin{align}\label{2.1}
\mathbb{W}^{\rm free}_{++}   = (v^+ \pa_+)^2 \int \frac{du}{(v^+u^+)^2} \mathbb{A}^{++}(u) \,.
\end{align}
Then we evaluate the amputated propagator
\begin{align}\label{WAamp}
\lim_{p^2\to0} p^2 \vev{\mathbb{W}^{\rm free}_{++}(p)  \mathbb{A}^{++}(-p)}\,.
\end{align}
Notice that this time the second end of the propagator is not a curvature as in \p{2.4} but the gauge prepotential itself. The reason is that our composite operators are made from supercurvatures which in turn are made from the prepotentials $A^{++},A^+_\da$, see \p{multbQ} and \p{ncW}. 
At this stage the odd variables $\q$ get projected with the negative helicity spinors $\la$, as explained in \p{3.7}. Then we do the half-FT \p{FT}, which eliminates the RHs $w$ from the external on-shell state. Notice that the free super-curvature  in \p{2.1} is in fact independent of the LH $v$, since it satisfies the constraint $\pa^{++}_v \mathbb{W}^{\rm free}_{++}=0$ \footnote{ The covariant counterpart of this relation takes the form $\pa^{++} \mathbb{W}_{AB} +g [\mathbb{A}^{++},\mathbb{W}_{AB}] = 0$ (recall \p{314}).}. The result is the supersymmetric analog of the state \p{ampu}. It  will be subsequently used in Sect.~\ref{MHV} for the calculation of MHV form factors by a substitution rule which is the analog of \p{subst}.

\subsection{Amputated chiral super-propagator}

 Let us first compute the chiral analog of \p{WAamp}, the amputated  propagator of the chiral truncation $W^{\rm free}_{++}(\q)$ of the super-curvature with the chiral prepotential $A^{++}$. The amputation of the external legs described in Sect.~2 requires a pole $1/p^2$ in the propagator. Our LHC propagator $\vev{A^{++} A^{++}}$ has such a pole, as we show below.
 
 We start by computing the propagator $\vev{W^{\rm free}_{++}A^{++}}$ with the help of \p{prop1}:
 \begin{align}\label{2.5'}
&\vev{W^{\rm free}_{++}(p,\q_+,v,w) A^{++}(-p,0,u)}  = (v^+ \pa_+)^2 \int \frac{du_1}{(v^+ u_1^+)^2} \vev{A^{++}(u_1) A^{++}(u)}\nt
&= (v^+ \pa_+)^2 \int \frac{du}{(v^+ u_1^+)^2} 4\pi \delta^2(p^{\-} u^+)\delta(u_1,u) \delta^4(\q\cdot u^+)
 = 4\pi \delta^2(p^{\-}u^+) \delta^2(w_{+A}^{a}\q^{A\a} u^+_{\a})\,.
\end{align}
The  differentiation $(v^+ \pa_{+})^2$  has been done by splitting 
$\delta^4(\q\cdot u^+)$ with the help of RHs, 
\begin{align}
(v^+ \pa_{+})^2\delta^4(\q\cdot u^+)   = (v^+ \pa_{+})^2[\delta^2(w_{-}\q u^+) \delta^2(w_{+} \q u^+) ]  = (v^+ u^+)^2 \delta^2(w_{+}\q u^+) \,.
\end{align}
We recall that in \p{2.5'} $p^{\-\a} = \xi^{\-}_{\da}p^{\da\a}$  is the  projection of the momentum with the light-cone gauge-fixing parameter  $ \xi^{\-}_{\da}$\,.    As expected, the $W$ end of the propagator  does not depend on the LH $v$ at that point, and it depends polynomially on the RH  $w$.

This propagator has a pole $1/p^2$. To reveal it, we recall \cite{Chicherin:2016fac} that { the bosonic delta function in \p{2.5'} identifies }
\begin{align}\label{2.7}
u^+_\a =p^{\-}_\a/\sqrt{p^2}\,, \qquad u^-_\a = - p^{\+}_\a/\sqrt{p^2}\,.
\end{align}
The LH $u$ will be integrated over, in expressions of the type
\begin{align}\label{2.8}
\int du \ \delta^2(p^{\-}u^+) \frac{P(u^+)}{Q(u^+)} = \frac1{\pi p^2}  \frac{P(p^{\-}/\sqrt{p^2})}{Q(p^{\-}/\sqrt{p^2})} 
= \frac1{\pi p^2}  \frac{P(p^{\-})}{Q(p^{\-})} \,.
\end{align}
Here $P,Q$ are homogeneous polynomials in $u^+$ of the same degree, so that their ratio has LH charge zero. Thus, after the integration the LH $u^+$ gets replaced by the projected momentum $p^{\-}$. The presence of a pole allows us to do the amputation.   On shell $p_{\a\da}= \la_\a \tl_\da$, so $p^{\-}_\a = [\xi^{\-}\tl] \la_\a$. Once again, due to the vanishing LH charge we can drop the factor $ [\xi^{\-}\tl]$. Thus, effectively (see \p{Wonshell})
\begin{align}\label{2.9'}
&\lim_{p^2\to0} p^2 \vev{W^{\rm free}_{++}(\q_+,w) A^{++}(\q_0,u)} \ \Rightarrow\nt
&\    \vev{\Phi_{++}(\q_+,w) A^{++}(\q_0,u)} =  \delta^2(w_{+}(\q-\q_0) \ket{\la}) \,  \delta(\la,u)\,, 
\end{align}
where the delta function can be treated as a harmonic one \p{delta}, identifying $u^+=\la$. We have also restored the dependence on $\q_0$ by translation invariance.  

What remains to do is to FT the Grassmann variable $\chi_+ = \vev{\q_+\la}$ (recall \p{3.7}) at the external leg as in \p{FT},
\begin{align}\label{2.10}
\int d^2\chi_+\ e^{\eta_-\chi_+ } \delta^2(\chi_+ - w_{+}\q_0 \ket{\la}) = e^{\eta_- w_{+}\q_0 \ket{\la}}\,.
\end{align}
So, finally, the on-shell state reads
\begin{align}\label{2.11}
\vev{\Phi(\eta_-) A^{++}(\q_0)} = \delta(\la,u)  e^{\eta_- w_{+}\q_0 \ket{\la}}\,.
\end{align}
This result is intermediate, we still need to restore the dependence on the other half $\eta_+$ of the odd variables to complete the on-shell state.

\subsection{The complete nonchiral on-shell state}

In the previous subsection we used the chiral truncation of the supercurvature $W^{\rm free}_{++}(x,\q_{+})$  in the analytic frame. In Sect.~\ref{s5.2} we reconstructed the full nonchiral supercurvature $\mathbb{W}^{\rm free}_{++}(x,\q_+,\bq_+)$ by working out the $\bQ$-variations of the chiral supercurvature. Schematically,
\begin{align}\label{3.1}
\mathbb{W}^{\rm free}_{++}(x,\q_+,\bq_+,u,w) = W^{\rm free}_{++}(\q_+) +  \ti \bq^\da_{+a'} \bQ^{a'}_{\da -} W^{\rm free}_{++} - \frac1{2} (\bq_+ \bQ_-)^2 W^{\rm free}_{++} + \ldots
\end{align}
We recall that the free supercurvature is manifestly R-analytic, i.e. it is annihilated by half of the super-charges, $\bQ_+ W^{\rm free}_{++}=0$. This is why we only use $\bQ_-W^{\rm free}_{++}$ in the expansion \p{3.1}.

The on-shell (amputated) $\bQ$-variation of the propagator $\vev{W^{\rm free}_{++} A^{++}}$ are given by (see \p{A.8})
\begin{align}\label{6.11}
&\lim_{p^2\to0} p^2 \vev{\bar Q_{-\da}^{a'}W^{\rm free}_{++}(\q_+,w) A^{++}(\q_0,u)}  = (-\ti) \tl_\da \, w_{-}^{a'} \vev{\q_0 \la}) 
\vev{\Phi_{++} A^{++}} \\
&\lim_{p^2\to0} p^2 \vev{\bar Q_{-\da}^{a'} \bar Q_{-\db}^{b'} W^{\rm free}_{++}(\q_+,w) A^{++}(\q_0,u)}  
= (-\ti)^2 \tl_\da  \tl_\db    (w_{-}^{a'}  \vev{\q_0 \la})   (w_{-}^{b'} \vev{\q_0 \la}) \vev{\Phi_{++} A^{++}} \label{6.12}
\end{align}
with $\vev{\Phi_{++} A^{++}}$ from \p{2.9'}.

Substituting these variations in  the antichiral expansion \p{3.1}, we see that the odd coordinates $\bq^\da$ get projected with the momentum helicity spinor $\tl_\da$, hence the expansion goes in the effective odd variables $\eta_{a'+} = \bq^\da_{a'+}\tl_\da$ (recall \p{3.7}). Then we find
\begin{align}\label{}
 \vev{\Phi_{++}(\q,\bq,w) A^{++}(\q_0,u)}  &=  [1+ \eta_{+}  w_{-}\q_0 \ket{\la} + \frac1{2} (\eta_{+}  w_{-}\q_0 \ket{\la})^2 ] \vev{\Phi_{++} A^{++}}\nt
&  =  e^{ \eta_{+}  w_{-}\q_0 \ket{\la}}\delta^2(w_{+}(\q-\q_0) \ket{\la}) \,  \delta(\la,u)\,.
\end{align}
The FT with respect to $\chi_{+} = \vev{\q_{+} \la} $ is performed as in \p{2.10}, resulting in  (from here on we drop the index $0$ at the $A^{++}$ end)
\begin{align}\label{3.5}
  \vev{\Phi(\eta) A^{++}(\q,u)} = \delta(\la,u)  e^{(\eta_{+}  w_{-} +\eta_- w_{+})\q \ket{\la}}  =  \delta(\la,u) e^{\eta\cdot \vev{\q  \la}}\,,
\end{align}
where we have used the completeness identity \p{5.1} for the RH $w$. 
As expected, the RH $w$ at the external end of the propagator has dropped out from the on-shell state.

Notice that the presence of the positive helicity spinor $\tl_\da$ in \p{6.11} explains why the on-shell anti-gluino becomes $\bar\psi_{\da} = \tilde\lambda_{\da} \bar\psi^{(-\frac1{2})}$ (cf. \p{states}). Similarly, the factor $ \tl_\da  \tl_\db$  in \p{6.12} explains why the negative helicity on-shell gluon is represented by $\tilde F_{\da\db} = \tilde\lambda_{\da}\tilde\lambda_{\db} g^{(-1)}$.

What we still need to do is to restore the $\bq$ dependence at the $A^{++}$ end of the propagator, $A^{++} \to \mathbb{A}^{++}$ (see \p{5.7}). Using the complete set of $\bQ-$variations from \p{A.8} and repeating the above steps, we find that the variable $\eta$ in the exponential in \p{3.5} is replaced by $\eta \ \to \ \eta + [\tl \bq]$. In this way find the nonchiral on-shell state
\begin{align}\label{6.14}
 \vev{\Phi(\eta) \mathbb{A}^{++}(\la,\tl,\q,\bq,u)}    =  \delta(\la,u) e^{ (\eta + [\tl\bq])  \vev{\q\la}}\,.
\end{align} 
According to \p{subst}, we need to  complete  the on-shell state by the space-time factor $e^{\ti x^\mu p_\mu } =e^{\ti/2 [\tl|x\ket{\la}}$. So, \p{6.14} gives rise to the substitution rule 
\begin{align}\label{6.15}
\mathbb{A}^{++}(\la,\tl,\q,\bq,u) \ \Rightarrow \ \delta(\la,u)\exp\left\{ \frac{\ti}{2} \tl_\da \left(  x^{\da\a} - 2\ti   \bq^\da_A \, \q^{\a A}  \right)\la_\a  +  \eta_A  \q^{\a A} \la_\a  \right\}
\end{align}
for each $\mathbb{A}^{++}$ in a composite operator made from these prepotentials. The applications  of this rule will be discussed in detail in Sect.~\ref{MHV}.

We remark that the odd variables $\q,\bq$ appear in \p{6.15} projected with helicity spinors, as claimed in \p{3.7}. 

Our result \p{6.15} is the nonchiral  generalization of the on-shell chiral state found in \cite{Adamo:2011cb}. The approach of Ref.~\cite{Adamo:2011cb} is to start from the collection of component states in the Wess-Zumino gauge and perform a gauge transformation to the CSW gauge. Here we have given a direct derivation of the on-shell state from first principles. We have also shown explicitly how the on-shell state becomes independent of the gauge-fixing parameter $\xi$.

\subsection{The on-shell state and supersymmetry}

Above we have derived the complete on-shell state by applying the amputation procedure to a supersymmetrized propagator. Here we wish to show that the main part of \p{6.15} can in fact be obtained by requiring invariance under  the $\cN=4$ supersymmetry algebra with generators
\begin{align}\label{6.16}
&Q_{\a A}= \ti\frac{\pa}{\pa \q^{\a A}} +2 \bq^{\da}_A \frac{\pa}{\pa x^{\da\a}} + \ti \la_\a \eta_A\,,   \qquad   \bQ^A_\da = -\ti\frac{\pa}{\pa \bq_A^\da} -2\q^{\a A} \frac{\pa}{\pa x^{\da\a}} +2\ti \tl_\da \frac{\pa}{\pa \eta_A} \nt
&\{Q_{\a A}, \bQ^B_\da\}=-2\delta_A^B \left( 2 i \frac{\pa}{\pa x^{\da\a}} + \la_\a \tl_\da \right) = -2\delta_A^B P_{\a\da} \,.
\end{align}

Indeed, let us start with the bosonic state $\exp\left\{{\ti}/{2}\,  \tl_\da x^{\da\a} \la_\a \right\}$. It is easy to see that this is the unique Lorentz invariant solution of the translation Ward identity $P f(x,p)=0$ with an on-shell momentum $p_{\a\da} =\la_\a \tl_\da$. Further, adding to the bosonic variables $x,p$ the odd variables $\eta,\q,\bq$ in all allowed Lorentz and dilation invariant combinations, one can show that the unique solution of the supersymmetry Ward identities $Qf(x,p,\eta,\q,\bq) = \bQ f(x,p,\eta,\q,\bq) = 0$ is indeed the exponential factor in \p{6.15}. 

Remarkably, the combination 
\begin{align}\label{}
x^{\da\a}_{\rm ch}=x^{\da\a} -2 \ti   \bq^\da_A \, \q^{\a A} 
\end{align}
which appears in \p{6.15}, has the meaning of a basis shift from the real superspace with space-time coordinate $x$ to the (left-handed) chiral superspace with coordinate $x_{\rm ch}$. The latter transforms as follows:
\begin{align}\label{CB}
Q_{\a A} x^{\db\b}_{\rm ch} =0\,, \qquad \bQ^A_\da x^{\db\b}_{\rm ch} = -4\delta_\da^\db \q^{\b A}\,,
\end{align}
i.e. it is inert under $Q-$supersymmetry. This result is not surprising, because  $\Phi$  in \p{6.14} has been constructed as a {\it chiral on-shell state}, see \p{css}. So, we can replace \p{6.15} by
\begin{align}\label{ChSub}
\mathbb{A}^{++}(\la,\tl,\q,\bq,u) \ \Rightarrow \ \delta(\la,u)\exp\left\{ \frac{\ti}{2} \tl_\da   x^{\da\a}_{\rm ch} \la_\a  +  \eta_A  \q^{\a A} \la_\a  \right\}\,.
\end{align}

In what follows we will make use of another basis, adapted to R-analytic superfields like the half-BPS operators \p{halfBPS}, in particular the stress-tensor multiplet \p{7.1}:
\begin{align}\label{AB}
x^{\da\a}_{\rm an}=x^{\da\a} +2 \ti   (\bq^\da_+ \,  \q^\a_- - \bq^\da_- \, \q^{\a}_+)\,,
\end{align}
where the odd variables are projected with RHs. 
In this basis  the spinor derivatives $D_+ \to \pa_+\,, \ \bar D_+ \to \bar\pa_+$ become short. Consequently, the R-analyticity  property  (independence of $\q_-,\bq_-$) of the  half-BPS operator $T$ becomes manifest, see \p{7.1}.

The relevance of the correct choice of basis in superspace becomes clear when we put a curvature at the second end of the amputated propagator  \p{3.5}, namely, $\vev{\Phi(\eta) \mathbb{W}^{\rm free}_{AB}}$. Now the `naive' definition \p{ncW} has to be modified. Instead of partial spinor derivatives $\pa^\a_A$ we have to use covariant ones. The latter are defined as operators anticommuting with the supersymmetry generators \p{6.16}:
\begin{align}\label{6.22}
&D_{\a A}= \frac{\pa}{\pa \q^{\a A}} +2\ti \bq^{\da}_A \frac{\pa}{\pa x^{\da\a}} \,,   \qquad   \bar D^A_\da = -\frac{\pa}{\pa \bq_A^\da} -2\ti\q^{\a A} \frac{\pa}{\pa x^{\da\a}}  \nt
&\{Q, D\}=\{\bQ, D\}= \{Q, \bar D\}=\{\bQ, D\}=0\,.
\end{align}
So, we need to compute (recall \p{ncA--}, \p{6.14})
\begin{align}\label{6.23}
&e^{\frac{\ti}{2} [\tl|x\ket{\la}} \vev{\Phi \mathbb{W}^{\rm free}_{AB}} =D^+_A D^+_B\int \frac{dv}{(u^+v^+)^2} \ e^{\frac{\ti}{2} [\tl|x\ket{\la}} \vev{\Phi   \mathbb{A}^{++}(v)} \nt
& =\frac1{\vev{u^+\la}^2} D^+_A D^+_B\  e^{ \frac{\ti}{2} [\tl|x\ket{\la} + (\eta+[\tl \bq])\vev{\q\la}  }= \hat\eta_A \hat\eta_B \  e^{ \frac{\ti}{2} [\tl|x\ket{\la} + (\eta+[\tl \bq])\vev{\q\la}  }\,,
\end{align}
where we see the $\bQ-$invariant combination
\begin{align}\label{7.4}
\hat\eta_{A} = \eta_{A} + 2[\tl\bq_A]\,, \qquad \bQ^B_\db \hat\eta_{A}=0\,.
\end{align} Here it was important to use the correct covariant derivatives $D^+_A$, with the super-torsion term $\bq\pa_x$, in order to obtain the $\bQ-$invariant \p{7.4}.

\subsection{MHV amplitude} 

As a very simple illustration, let us apply our substitution  rule \p{ChSub} to the $n-$point MHV super-amplitude.  The  Zupnik (interaction) term in the $\cN=4$ SYM action has a form similar to \p{448} (see \cite{Chicherin:2016fac}),
\begin{align}\label{Zup}
S_Z =  \tr\sum^\infty_{n=2} \frac{(-g)^{n-2}}{n}\int d^4x d^8\q\  du_1\ldots du_n\; { 
A^{++}(x,\q \cdot u^+_1,u_1) \ldots A^{++}(x,\q \cdot u^+_n,u_n) \over(u^+_1u^+_2) \ldots
(u^+_nu^+_1)}\,.
\end{align} 
Here the bilinear term in fact belongs to the free action, so the true interaction terms have $n\geq3$. Now, consider the $n-$valent Zupnik vertex and substitute each chiral $A^{++}$ by the chiral  on-shell state \p{ChSub}. The delta functions remove the LH integrals and replace the LHs by negative helicity spinors $\la$.  The exponentials from \p{ChSub}, after the integration over the vertex point $\int d^4x \, d^8 \q$, produce the complete super-momentum conservation delta function and we recover the familiar result for the $n-$particle amplitude \cite{Nair:1988bq}
\begin{align}\label{}
A^{\rm MHV}_n = \frac{\delta^4(\sum \la_i \tl_i) \delta^8(\sum \la_i \eta_i)}{\vev{12} \ldots \vev{n1}}\,.
\end{align}

\section{MHV form factors}\label{MHV}

The construction of the full on-shell state \p{6.15} has led to the following very simple substitution rule for the computation of MHV form factors. Consider an operator in the form of a product of supercurvatures $\mathbb{W}_{AB}$ (or their derivatives, for details see Sect.~\ref{allop}).  Each $\mathbb{W}_{AB}$ is made from many $\mathbb{A}^{++}$, see \p{ncW}. In the form factor each $\mathbb{A}^{++}$ from a given vertex is connected to an external state by a free propagator.  The substitution rule is to replace the $i-$th external leg at the vertex by the super-state \p{6.15}. The bosonic delta function  $\delta(\la_i,u_i)$ removes the LH integral at the vertex and replaces the LH $u^+_i$ by the negative helicity spinor $\la_i$. 

In this section we consider in detail two simple examples of the application of this rule, the form factors of the stress-tensor and the Konishi multiplets.

\subsection{The stress-tensor multiplet}

The $\cN=4$ SYM stress-tensor multiplet is the simplest of the half-BPS operators defined in \p{halfBPS}:
\begin{align}\label{7.1}
T =\tr (\mathbb{W}_{++})^2(x_{\rm an},\q_+,\bq_+,w)\,,
\end{align}
where the R-analytic basis coordinate $x_{\rm an}$ was defined in \p{AB}. 
We wish to evaluate its tree-level MHV form factor. 
To this end we need to first compute the Born-level correlation function
\begin{align}\label{4.3}
\vev{\mathbb{W}^{\rm free}_{++}(1) \ldots \mathbb{W}^{\rm free}_{++}(n) \tr (\mathbb{W}_{++})^2(x_{\rm an},\q_+,\bq_+,w)}\,,
\end{align}
then amputate the external legs $1,\ldots,n$.  We consider the color ordered part of \p{4.3}, which implies cyclic ordering of the external legs.

\begin{figure}
\centering
\includegraphics[height = 3cm]{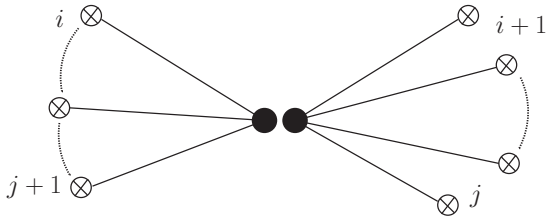}
\label{fig1}
\caption{The external legs are separated into two clusters 
$i+1 \leq l \leq j$ and $j+1 \leq l \leq i$. They are linked by the amputated propagators \p{6.15} 
with the supercurvatures $\mathbb{W}_{++}$ in  the composite operator $T$ \p{7.1}.}
\end{figure}

The supercurvature $\mathbb{W}_{++}$ is defined in  \p{ncW}, \p{W++}. The Born-level correlation function \p{4.3} is obtained by connecting each external leg $\mathbb{W}^{\rm free}_{++}(l)$ with an $\mathbb{A}^{++}$ inside the composite operator $\tr (\mathbb{W}_{++})^2$ by a free propagator. The resulting expression is proportional to $g^{n-2}$, as expected from the tree-level form factor. This procedure splits the correlator in two clusters. Let us choose two legs with labels $i<j$. The first cluster contains the legs with $i+1 \leq l \leq j$, the second contains the legs with $j+1 \leq l \leq i$ (the legs are labeled in a cyclic way). The legs from the first cluster are contracted with the first factor $\mathbb{W}$ from the composite operator, the  remaining  legs are contracted with the second factor $\mathbb{W}$. The complete correlator is obtained by summing over all values of $1 \leq i<j \leq n$ (see Fig. \ref{fig1}).

Let us examine the structure of a cluster made of $k$ legs. After the amputation, the calculation amounts to applying the substitution rule \p{6.15} to each leg within a given cluster. The bosonic delta functions $\delta(\la_l, u_l)$ remove the integrals in \p{ncA--} and replace the LHs by the negative helicity spinors. The calculation basically repeats \p{6.23}.  The result for a cluster of $k$ legs is simply
\begin{align}\label{4.6}
& \Gamma(1,2,\ldots,k)\equiv \vev{\prod_{l = 1}^{k}  \, \Phi(p_l,\eta_l) \cdot
\mathbb{W}_{++}(x_{\rm an},\q,\bq,w,u)|0}_{\rm tree\ MHV} \nt
&  =	(u^+D_{+})^2 \frac{ e^{\frac{\ti}{2}  \sum_{l = 1}^k z_l}}{\vev{u^+ 1} \vev{12} \ldots \vev{k u^+}}  = \left( \sum_{l = 1}^k \vev{u^+ l}  \hat\eta_{l +} \right)^2
\frac{  e^{  \frac{\ti}{2} \sum_{l = 1}^k z_l}}{\vev{u^+ 1} \vev{12} \ldots \vev{k u^+}}\,,  
\end{align} 
where $z_l=[l|(x_{\rm an}-4\ti\bq_+\q_-)\ket{l} -2 \ti \eta_{l}\vev{\q l}$.  
Each $\hat\eta_{l a'+} = \bar w^A_{a'+} \hat\eta_{l A}$ is projected with the  RH of the composite operator and we have made the appropriate change of basis.  

We remark that this expression still depends on the LH $u$ of the supercurvature. This dependence is due to the analytic gauge frame.   It disappears once the gauge invariant composite operator has been reconstructed by combining together the two clusters: 
\begin{align}\label{4.7}
&\vev{1, 2, \ldots, n |T(x_{\rm an},\q_+,\bq_+,w)|0 }_{\rm tree\ MHV}  = \sum_{i<j} \Gamma(i+1,\ldots,j)\Gamma(j+1,\ldots,i)\nt 
&  = \frac{1}{\vev{12}\ldots \vev{n1}} \ \delta^4\left( \sum_{l=1}^n  \hat\eta_{l +} \ket{l} \right)\   
\exp  \left\{\sum_{l = 1}^n \left({\textstyle\frac{\ti}{2}} [l|x_{\rm an}\ket{l} +  \eta_{l-} \vev{\q_+ l} \right) \right\}\,,
\end{align}  
 where $\hat\eta$ was defined in \p{7.4}. 
In deriving this expression we have applied the eikonal identity \cite{Dixon:1996wi}
\begin{align}\label{eikonal}
\sum_{k=i}^{j-1} \frac{\vev{k\, k+1}}{\vev{k u^+} \vev{u^+\, k+1}} = \frac{\vev{i j}}{\vev{iu^+} \vev{u^+\,j}} 
\end{align}
 twice. This identity is responsible for the elimination of the LH $u^+-$dependence  at the operator point. 

Notice that the final expression depends only on the RH projected odd variables $\q_+,\bq_+$, as should be for the half-BPS operator $T$ \p{7.1}.   It  is annihilated by  the generators of  supersymmetry  \p{6.16} adapted to the R-analytic basis \p{AB}.

Our result coincides with that of  \cite{Brandhuber:2011tv} where the form factor has not been computed but rather predicted from supersymmetry Ward identities. Here we have explained how the direct computation using our LHC field theory rules leads to the desired result. 

We can obtain the same result more quickly if we use the alternative representation 
of the chiral truncation of the stress-tensor multiplet in terms of the interaction (Zupnik) term of the Lagrangian, $\cT = (\pa_+)^4 L_{\rm Z}$ (see \cite{Chicherin:2014uca}). The Lagrangian $L_{\rm Z}$ can be read off from the Zupnik action \p{Zup},  $S_{\rm Z} = \int d^4 x d^8 \q L_{\rm Z}$. Doing the substitution \p{3.5} we obtain
\begin{align}
\vev{1, 2, \ldots, n | \mathcal{T}(x,\q_+,w)|0 }_{\rm tree\ MHV}& = \frac{1}{\vev{12}\ldots \vev{n1}}
 (\pa_{+})^4  e^{\sum_{l = 1}^n \left(\frac{\ti}{2}[l|x\ket{l}+ \eta_{l} \vev{\q l}\right)} 
 \end{align}
which immediately gives the chiral truncation of \p{4.7}. Then the $\bq-$dependence can be restored by supersymmetry. Notice that in this derivation we do not need the eikonal identity \p{eikonal}. However, this shortcut is only possible for the stress-tensor multiplet. The other operators have to be constructed in terms of the supercurvature $W_{AB}$, for example, the Konishi multiplet \p{konishi}.

\subsection{Konishi multiplet}

We consider the full Konishi multiplet $K(x,\q,\bq) = \ep^{ABCD} \tr \left( \mathbb{W}_{AB} \mathbb{W}_{CD} \right)$, i.e. the nonchiral extension of the chiral truncation \p{konishi}. 
The calculation of the MHV tree level form factor for $K$ follows the pattern of the previous subsection.
Firstly, we examine the contribution of a cluster containing $k$ external legs which are contracted by the amputated propagators 
with the nonchiral supercurvature $\mathbb{W}_{AB}$ \p{ncW}. The substitution rule \p{6.15} gives rise to (recall \p{6.23})
\begin{align}\label{}
& \Gamma_{AB}(1,2,\ldots,k)\equiv \vev{\prod_{l = 1}^{k}  \, \Phi(p_l,\eta_l) \cdot
\mathbb{W}_{AB}(x,\q,\bq,w,u)|0}_{\rm tree\ MHV} \nt
&  =	\left( \sum_{l = 1}^k \vev{u^+ l} \hat\eta_{l A}  \right)
\left( \sum_{l = 1}^k \vev{u^+ l} \hat\eta_{l B}  \right)
\frac{  e^{  \frac{\ti}{2} \sum_{l = 1}^k z_l}}{\vev{u^+ 1} \vev{12} \ldots \vev{k u^+}}\,,  
\end{align} 
where $z_l=[l|(x-2\ti\bq\q)\ket{l} -2 \ti \eta_{l}\vev{\q l}$ and $\hat\eta$ was defined in \p{7.4}. This amputated correlation function is gauge covariant, but not gauge-invariant. Being defined in the analytic gauge frame,  it depends on the LH $u$. 
This dependence disappears in the gauge-invariant form factor, again due to the eikonal identity \p{eikonal}  
\begin{align}\label{}
&\vev{1, 2, \ldots, n |K(x,\q,\bq,w)|0 }_{\rm tree\ MHV}  = \ep^{ABCD}\sum_{i<j} \Gamma_{AB}(i+1,\ldots,j)\Gamma_{CD}(j+1,\ldots,i)\nt 
&  = \frac{e^{  \sum_{l = 1}^n \frac{\ti}{2}[l|(x - 2\ti \bq\q)\ket{l} + \eta_l \vev{\q l} }}{\vev{12}\ldots \vev{n1}} \ \sum_{i\leq j < k \leq l}  
(2-\delta_{ij})(2-\delta_{kl})\, \ep^{ABCD} \hat\eta_{i A}\hat\eta_{j B}\hat\eta_{k C}\hat\eta_{l D} \, \vev{jk} \vev{li} \,.
\end{align}  
Notice that  the odd factor of degree 4 does not form a fermionic delta function, unlike the stress-tensor multiplet in \p{4.7}.

\section{Component operators} \label{allop}

In this Section we construct the composite operators in $\cN = 4$ SYM 
out of the scalar and (anti)-gluino fields, YM curvatures and YM covariant derivatives,  
providing their formulation in terms of L-analytic superfields (see Sect.~\ref{LHC}). 
We prefer not to start with the most general operators. Instead we first consider  
a series of examples for some subclasses of operators. 
We begin with a subclass admitting the simplest LH formulation
and complete the construction by the most general operators.
In some cases the construction is rather involved. Nevertheless, 
we will see in Sect.~\ref{compFF} that the MHV tree-level form factors are blind to many of these subtle details.

\subsection{Lowest twist operators}

Instead of working with multiplets of operators like in Sect.~\ref{opers}, we can also construct their components one by one.
To this end we need to extract component fields from the supercurvature $W_{AB}$. 
Taking  derivatives  $\pa^+_A$, which do not have a gauge connection in the analytic frame (see Sect.~\ref{LHC}), 
we define the LH fields
\begin{align}
&\phi_{AB}(x,u) \equiv W_{AB}(x,\q,u)|_{\q = 0} \nt
&\psi^{+A}(x,u) \equiv 1/3!\, \ep^{ABCD} \pa^+_B W_{CD}(x,\q,u)|_{\q = 0} \nt
&F^{++}(x,u) \equiv 1/4!\, \ep^{ABCD} \pa^{+}_{A} \pa^{+}_{B} W_{CD}(x,\q,u)|_{\q=0} \label{Wcfield}
\end{align}
with $U(1)$ charges  $0$, $(+1)$, $(+2)$, respectively. 
They transform covariantly under gauge transformations with an LH field parameter $\Lambda(x,u)$ 
since we have set $\q \to 0$ (recall \p{314}).
 Eqs.~\p{Wcfield} take the 
following form in terms of the L-analytic prepotential $A^{++}$ (recall \p{448})
\begin{align}
&\phi_{AB}(x,u) = \pa^+_{A} \pa^+_{B} A^{--}|_{\q =0} \ \ , \ \
\psi^{+A}(x,u) = (\pa^+)^{3 A}  A^{--}|_{\q = 0} \ \ , \ \ F^{++}(x,u) = (\pa^{+})^4 A^{--}|_{\q=0} \notag
\end{align}
Multiplying together  several LH fields from \p{Wcfield} and taking the trace, we obtain a gauge invariant operator.\footnote{As mentioned earlier, in this paper we consider only single trace operators.}
For example, 
\begin{align}
\cO^{+++A}(x,u) = \tr \left( F^{++} \psi^{+A} \right) \,.
\end{align}
Unlike the Konishi multiplet \p{konishi},
this operator depends on the LHs in a polynomial way, 
$\cO^{+++A}(x,u) = u^{+\a} u^{+\b} u^{+\gamma}\cO_{\a\b\gamma}^{A}(x)$. 
To see this,  we profit again from lemma \p{5}. From $\nabla^{++} W_{AB} = 0$ (recall \p{314}) and $[\nabla^{++},\pa^+_A] = 0$ 
(L-analyticity of $A^{++}$) it follows that $\pa^{++}\cO^{+++A}(x,u)  = 0$. Then, inside the gauge invariant operator we can identify the LH 
fields \p{Wcfield} with the physical fields, 
\begin{align}
\cO_{\a\b\gamma}^{A}(x) = \tr \left( F_{(\a\b} \psi_{\gamma)}^{A}\right).
\end{align}

We can also easily include covariant YM derivatives $\nabla_{\a\da} = \pa_{\a\da} + g\cA_{\a\da}$ in the game. In the analytic frame it corresponds to acting with several covariant derivatives 
$\nabla^+_{\da} = \pa^+_{\da} + g A^+_{\da}$ on the LH fields \p{Wcfield} and then setting $\q \to 0$. 
To obtain irreducible representations we symmetrize the dotted Lorentz indices.
Then we form the product of several fields as before. The polynomial dependence on the LHs
of the gauge-invariant operators follows again from lemma \p{5} and the equation of motion of $\cN = 4$ SYM, 
$[\nabla^{++},\nabla^+_{\da}] = 0$ \cite{Chicherin:2016fac}. 
Eliminating the LHs we find the corresponding gauge-invariant operator in terms of the physical fields.
For example,
\begin{align}
\tr\left( \nabla^+_{(\da} \nabla^+_{\db)} \phi_{AB} \nabla^+_{\dot\gamma}\psi^{+C} \right)(x,u) \ \leftrightarrow \
u^+_{\a} u^{+}_{\b} u^+_{\gamma} u^+_{\delta} \,\tr\left( \nabla^{\a}_{(\da} \nabla^{\b}_{\db)} \phi_{AB} 
\nabla^{\gamma}_{\dot\gamma}\psi^{\delta C} \right)(x) \,.
\end{align}
Let us emphasize that we have to use covariant derivatives with the gauge connection $A^+_{\da}$ to produce dotted Lorentz indices.

Thus, working in the analytic frame and using the chiral supercurvature $W_{AB}$ and the covariant derivatives $\pa^+_A$, $\nabla^+_{\da}$ 
we are able to construct gauge invariant operators made of the scalars $\phi_{AB}$, gluinos $\psi^{A}_{\a}$, self-dual YM curvatures $F_{\a\b}$,
and covariant YM derivatives $\nabla_{\a\da}$. All undotted Lorentz indices of the gauge invariant operator
are symmetrized, which corresponds to the lowest twist.

In order to include the anti-gluinos $\bar\psi^{\da}_{A}$ and the anti-self-dual YM curvatures $\tilde F_{\da\db}$
in the construction of composite operators we have to apply the $\bar Q$-transformations \p{bQ} to the chiral supercurvature.
We will need single and double $\bQ$-variations since these component fields are accompanied by one and two $\bq$, respectively 
(the second and the third line in \p{Wexp}). In terms of the nonchiral supercurvature \p{ncW}
\footnote{Here we need only self-dual part of $\bQ$-transformations \p{bQ}, i.e. 
we can throw away $\bQ_{\rm Z}$. Since $W_{AB}$ depends only on $A^{++}$, so $\bQ_{\rm Z}$ does not appear in the first $\bQ$-variation. 
It does not arises as well in the second $\bQ$-variation in \p{Wncfield} due to the index symmetrization $(\da\db)$ (see the Appendix). } 
\begin{align}
&\bar\psi_{\da A}(x,u) \equiv  \ti \bar Q^{B}_{\da} W_{AB}(x,\q,u)|_{\q = 0} 
= \bar \pa^{B}_{\da} \mathbb{W}_{AB}|_{\q =\bq = 0} \,,\nt
&\tilde F_{\da\db}(x,u) \equiv \ti^2 \bar Q^{A}_{\da} \bar Q^{B}_{\db} W_{AB}(x,\q,u)|_{\q=0} 
= \bar \pa^{A}_{\da} \bar \pa^{B}_{\db} \mathbb{W}_{AB}|_{\q =\bq = 0} \,,\label{Wncfield}
\end{align}
where $\bar \pa^{A}_{\da} \equiv \pa / \pa \bq^{\da}_{A}$.
The LH independence of the gauge invariant operators involving the LH fields \p{Wncfield} is proved by 
 lemma \p{5} and $[\nabla^{++},\bar Q^A_{\da}] \sim \q$ that follows from the $\bQ$-transformation law of 
$A^{++}$ \p{bQ}. The $\bQ$-transformations commute with the gauge transformations \cite{Chicherin:2016fac}, so 
the gauge covariance of $W_{AB}$ \p{314} is not spoiled by the $\bQ$-variations.

We collectively denote by $\mathcal{W}$ the LH fields from \p{Wcfield} and \p{Wncfield},
carrying a number of covariant derivatives $\nabla^+_{\da}$ and taken at $\q = 0$,
\begin{align}\label{calW}
\mathcal{W}(x,u) \ \ \leftrightarrow \ \ \nabla^+_{\da_1} \ldots \nabla^+_{\da_r} \
\{\ \phi_{AB}\ , \ \psi^{+A} \ ,\  F^{++}\ ,\ \bar\psi_{\da A}\ ,\ \tilde F_{\da\db}\ \}(x,u)\,.
\end{align}
We can rewrite it as well as number of $\pa^+_A$ and $\bar\pa^{A}_{\da}$ acting on $\mathbb{A}^{--}$ \p{ncA--},
\begin{align}\label{calW'}
\mathcal{W}(x,u) \ \leftrightarrow \ \nabla^+_{\da_1} \ldots \nabla^+_{\da_r} 
\{ \ \pa^+_{A}\pa^+_{B}\ , (\pa^{+})^{3 A} , (\pa^{+})^4 ,\ \bar\pa_{\da}^B\pa^+_A \pa^+_B ,\ \bar\pa_{\da}^A\bar\pa_{\db}^B \pa^+_A \pa^+_B\, \} 
\ \mathbb{A}^{--}|_{\q = \bq = 0},
\end{align}
where $(\pa^{+})^{3 A}$ and $(\pa^{+})^4$ are defined similar to \p{etapow}.
The composite gauge invariant operators are traces of their products\footnote{It is important to recall that in general such operators need diagonalization to become eigenstates of the dilatation operator (see, e.g., \cite{Belitsky:2003sh,Engelund:2012re}). We do not address this issue here.  }
\begin{align}\label{Ltw}
\tr \left( \mathcal{W}_1(x,u) \ldots \mathcal{W}_m(x,u) \right)\,.
\end{align}
The same  argument as above shows
that they are monomials in the LH $u^+$. This LH also symmetrizes the undotted 
Lorentz indices thus picking up the lowest twist. The LH fields transform covariantly under the gauge transformations
\begin{align}\label{calWgauge}
\mathcal{W}(x,u) \to e^{-g\Lambda(x,u)} \mathcal{W}(x,u) e^{g\Lambda(x,u)}\,.
\end{align}
Let us emphasize that both prepotentials $A^{++}$ and $A^+_{\da}$ appeare in $\mathcal{W}$,
and so are indispensable in the construction of composite operators.

\subsection{Central gauge frame} \label{centfr}

So far we have worked in the analytic frame, in which the gauge connections transform with an L-analytic parameter $\Lambda(x,\q^+,u)$
\begin{align} \label{analytGT}
\delta_{\Lambda} A = \nabla \Lambda \ \ \ \ \mbox{where $A$ denotes}\ \ A^{++}, A^{--}, A^{-}_A, A^{\pm}_{\da}\,.
\end{align}
As a consequence all the supercurvatures that we use  depend on the LHs in a non-polynomial way, 
even though the gauge invariant operators are { polynomials in  the LHs}. This property can be made manifest by switching to the so called central (or $\tau$-) frame \cite{Galperin:1984av,Mason:2005zm}. 
There the $SU(2)_L$ algebra  \p{su2} of the LH derivatives becomes flat but $\pa^+_{A}$ acquires a gauge connection instead,
\begin{align}\label{}
\{ \nabla^{++} , \nabla^{--} , \pa^0 , \pa^+_A , \nabla^-_A , \nabla^{\pm}_{\da} \}\ \xrightarrow{h}\ 
\{ \pa^{++} , \pa^{--} ,  \pa^0 , u^{+\a}\nabla_{\a A}  , u^{-\a}\nabla_{\a A} , u^{+\a}\nabla_{\a\da} \}\,.
\end{align}
The `bridge' relating the analytic and $\tau$-frames has the form of a generalized finite gauge transformation  $h(x,\q,u)$  \cite{Galperin:1984av,Galperin:1987wc}.
In particular, $h^{-1} \nabla^{++} h = \pa^{++}$ and $h^{-1} \pa^{+}_{A} h = u^{+\a}A_{\a A}(x)$.
In the $\tau$-frame the LH expansions like \p{3} reduce to a single polynomial term, so the LHs can be stripped off. 
In the $\tau$-frame the gauge transformations of the super-connections are $\delta_{\tau} A = \nabla \tau$ 
with a chiral LH-independent parameter $\tau = \tau(x,\q)$.
The bridge $h$ undergoes gauge transformation with respect to both the analytic and $\tau$-frames,
\begin{align} \label{gaugebridge}
h(x,\q,u) \ \rightarrow \  e^{-g\Lambda(x,\q^+,u)}h(x,\q,u) e^{g\tau(x,\q)}  \,.
\end{align}

The covariant LH-independence of the supercurvature $W_{AB}(x,\q,u)$ \p{314} in the analytic frame
is translated into the LH-independence of the supercurvature $W_{AB}(x,\q)$ in the $\tau$-frame,
\begin{align} \label{Wnoharm}
W_{AB}(x,\q) = h^{-1}(x,\q,u) W_{AB}(x,\q,u) h(x,\q,u)\,.
\end{align}
Thus the bridge $h$ effectively strips off the dependence on the LHs.
The elimination of the LHs via eq.~\p{Wnoharm} is an unnecessary step if one is interested in constructing gauge invariant objects out of $W_{AB}$. Indeed, the bridge transformation \p{Wnoharm} drops out from gauge invariant operators. 

In the $\tau$-frame the fields \p{Wcfield} (the constituents of the composite operators) take the form
\begin{align}
&\phi_{AB}(x) = W_{AB}(x,\q)|_{\q = 0} \ \ , \ \
\psi^{A}_{\a}(x) = 1/3!\, \ep^{ABCD} \nabla_{\a B} W_{CD}(x,\q)|_{\q = 0} \nt
&F^{\a\b}(x) = 1/4!\, \ep^{ABCD} \nabla^{\a}_{A} \nabla^{\b}_{B} W_{CD}(x,\q)|_{\q=0} \notag
\end{align}
where we have stripped off the LHs. So in the $\tau$-frame the {\it covariant }derivatives $\nabla^{\a}_A$, $\nabla^{\a}_{\da}$ are
indispensable for producing undotted Lorentz indices.

\subsection{Higher twist operators}

Above we have presented an LHC construction of operators with totally symmetrized undotted Lorentz indices.  
Indeed the undotted Lorentz indices arise from acting with $\pa^+_A$ and $\nabla^+_{\da}$ on 
the supercurvature $W_{AB}$ (recall \p{Wcfield}), taking $\q \to 0$, forming a gauge-invariant operator out of them, 
and then stripping off the LHs. So all undotted Lorentz indices are contracted with 
the same LH $u^{+}_{\a}$ and hence are symmetrized.

Now we want to construct operators where some of the Lorentz indices can be contracted, which corresponds to higher twist. 
Some higher twist operators live in the supermultiplets considered in Sect.~\ref{opers}.
There are two ways to avoid the automatic symmetrization of the undotted Lorentz indices. 

The first possibility is provided by the covariant derivatives $\nabla^{\a}_A$ and $\nabla^{\a}_{\da}$ 
written in the analytic frame
\begin{align}\label{911}
&\nabla_{\a A} = u^-_{\a} \pa^+_{A} - u^+_{\a} \nabla^{-}_{A} = \pa_{\a A} - u^+_{\a} g A^-_{A}\nt
&\nabla_{\a \da} = u^-_{\a} \nabla^+_{\da} - u^+_{\a} \nabla^{-}_{\da} = \pa_{\a\da}+ u^-_{\a} g A^{+}_{\da} - u^+_{\a} g A^{-}_{\da} \,.
\end{align}
We can use them  instead of $\pa^+_A$ and $\nabla^+_{\da}$ to produce Lorentz indices in \p{Wcfield}.
The gauge connections from \p{911} are expressed in terms of the 
prepotentials $A^{++}$, $A^+_{\da}$ as follows:
$A^-_{A} = - \pa^+_{A} A^{--}$, $A^-_{\da} = \pa^{--} A^{+}_{\da} - \pa^+_{\da} A^{--} + g [A^{--},A^+_{\da}]$ whith
$A^{--}$ given in \p{448}.
 
The second possibility is to consider each field $\cal W$ \p{calW} constituting the composite operator 
in its own analytic frame depending on its own LH $u$. Then the use of 
$\pa^+_A$ and $\nabla^+_{\da}$ implies the symmetrization of the undotted Lorentz indices of each constituent field.
If we work with several analytic frames,  we need bridge transformations relating them.
Combining a pair of $h$ bridges, we obtain the transformation from the analytic frame with LHs $v$
to the analytic frame with LHs $u$,  
\begin{align}\label{Uh}
U(x,\q;u,v) = h(x,\q,u) h(x,\q,v)^{-1}\,.
\end{align}
It is inert under the $\tau$-frame gauge transformations but transforms with respect to both
analytic frames (see \p{gaugebridge}),
\begin{align} \label{gb}
 U(x,\q;u,v) \ \rightarrow \  e^{-g\Lambda(x,\q^+,u)} U(x,\q;u,v) e^{g\Lambda(x,\q^+,v)}\,.
\end{align}
In the twistor literature \cite{Mason:2005zm,Mason:2010yk} this object is called a  `parallel propagator', 
and $U(u,v)$ is interpreted as a holomorphic Wilson line in \cite{Bullimore:2011ni}.
An explicit expression for $U$ in terms of the dynamical field $A^{++}$ 
can be found solving an LH differential equation with the boundary condition $U(u,u)=1$ (see \p{Uh}),
\begin{align} \label{U}
U(u,v) = 1 + \sum_{n = 1}^{\infty} (-g)^n \int d u_1 \ldots d u_n
\frac{(u^+ v^+)A^{++}(1)\ldots A^{++}(n)}{(u^+ u_1^+)(u_1^+ u_2^+) \ldots (u_n^+ v^+)} \,,
\end{align}
which is very similar to $A^{--}$ \p{448}. In fact $A^{--}$ appears in the Taylor 
expansion $U(u,v)= 1+ (u^+ v^+) g A^{--} +\ldots $ at $v \to u$ \cite{Lovelace:2010ev}.

Then we form the gauge-invariant operators out of the LH fields $\mathcal{W}_i$ (see \p{calW}),
each living in its own analytic frame specified by the LHs $u_i$, and we connect them by $U$-bridges taken at $\q \to 0$, 
\begin{align} \label{hloop}
\cO(x;u_1,\ldots,u_m) = \tr \left( \mathcal{W}_1(u_1) U(u_1,u_2) \mathcal{W}_2(u_2) U(u_2,u_3) \ldots  \mathcal{W}_m(u_m) U(u_m,u_1)\right)\,.
\end{align}
{We remark that this operator is nonlocal in the LH space but remains local in space-time.}
The gauge invariance of \p{hloop} follows from \p{calWgauge} and \p{gb}. 
The polynomiality of \p{hloop} with respect to $u_1 , \ldots, u_m$ can be seen by switching to the central frame (see Sect.~\ref{centfr}). 
The $U$-bridge \p{U} is constructed out of $A^{++}$, but $\mathcal{W}$ includes 
both prepotentials $A^{++}$ and $A^+_{\da}$.
{ In the next subsection we prefer this formulation for the calculation of form factors, since it facilitates the combinatorics.}

{ In conclusion, we repeat that this nonlocal construction of operators is only needed if we want to contract some of the undotted indices of the constituent fields. In all other case the use of the $U$-bridge is superfluous.  }

\subsection{Form factors} \label{compFF}

Having the composite operators formulated in terms of L-analytic superfields, 
the calculation of the MHV tree-level form factors is straightforward. We consider the form factor of the
the composite operator \p{hloop},
\begin{align} \label{hloopFF}
\vev{1,2,\ldots,n| \cO(x;u_1,\ldots,u_m) |0}^{\rm tree}_{\rm MHV}\,.
\end{align}
When the LHs at each leg coincide, $u_1 = \ldots = u_m \equiv u$,  we are dealing with the lowest twist operators \p{Ltw}.

The calculation follows the scheme from Sect.~\ref{MHV}. We consider a cluster of $k$ 
scattering states and a $\mathcal{W}$ (or $U$) constituting the operator \p{hloop}.
Then we substitute for each prepotential $\mathbb{A}^{++}$
inside each  $\mathcal{W}$ (or $U$)  the on-shell states \p{6.15}.
In the case of the $U$-bridge \p{U} the result is especially simple. Indeed the bridge is chiral and we take it at $\q = 0$, 
\begin{align}\label{Ucluster}
 \vev{\prod_{l = 1}^{k}   \, \Phi(p_l,\eta_{l}) \cdot
U(x;u,v) |0}^{\rm tree}_{\rm MHV} = \	g^{k} 
\frac{ \vev{u^+ v^+} e^{ \textup{i} x P}}{\vev{u^+ 1} \vev{12} \ldots \vev{k v^+}} 
\end{align}
where $P = p_1 + \ldots + p_k$.
Proceeding to the harmonic fields $\mathcal{W}$ in \p{hloop} 
we first note that the covariant derivatives $\nabla^+_{\da} = \pa^+_{\da} + g A^+_{\da}$
in \p{calW} effectively reduce to $\pa^+_{\da}$. The prepotential $A^+_{\da}$ does not contribute at the MHV tree level. 
Indeed, the propagator $\vev{A^+_{\da} A^{++}}$ \p{prop2} does not contain a pole $1/p^2$, 
so $\lim_{p^2 \to 0} p^2 \vev{\mathbb{W}(p,\eta) A^+_{\da}} = 0$. 
Thus the contribution of the cluster contracted with $\cal W$ takes the form  
\begin{align}\label{Wcluster}
 \vev{\prod_{l = 1}^{k}   \, \Phi(p_l,\eta_{l}) \cdot
{\cal W}(x,u) |0}^{\rm tree}_{\rm MHV} \ \leftrightarrow \	g^{k-1}
P^+_{\da_1} \ldots P^+_{\da_r} 
\frac{ e^{  \textup{i} x P + \sum_{l = 1}^k  (\eta_{l} + [l\bq])\vev{\q l}}}{\vev{u^+ 1} \vev{12} \ldots \vev{k u^+}} 
\end{align}
where $P^+_{\da} = u^{+\a} P_{\a\da}$. Here we are acting on the right-hand side of \p{Wcluster} 
with a number of derivatives  $\pa^+_A$ and $\bar\pa^{A}_{\da}$ and then setting $\q = \bq = 0$ to specify $\cal W$
according to \p{calW'}.

The form factor \p{hloopFF} is then obtained by multiplying together the contributions \p{Ucluster} and \p{Wcluster} of $2m$ clusters (see \p{hloop}), 
and by summing over all ways of distributing $n$ on-shell states among $2m$ clusters preserving the cyclic ordering.

We know that the composite operators \p{hloop} are polynomial in the LHs $u_1^+,\ldots,u^+_m$. This
is guaranteed by the invariance of \p{hloop} with respect to the gauge transformations in the analytic frame \p{analytGT}.
The contributions of the clusters \p{Ucluster} and \p{Wcluster} are not polynomial in the LHs.
The polynomiality applies only to the form factor \p{hloopFF}, which is a gauge invariant quantity,
after assembling the contributions of all clusters. The explicit elimination of the spurious poles in the LHs 
is a purely algebraic problem that can be easily solved applying $2m$ times the eikonal identity \p{eikonal}.
The combinatorics is the same as in \cite{Koster:2016loo}. This results in the formula (4.6) given there. 
{We do not reproduce this formula here. The purpose of our paper is to explain 
the field theory origin of the recipe given in \cite{Koster:2016loo} for the calculation of the MHV tree level form factors.}

\section{Conclusions}

In this paper we pursue two goals. Firstly, we formulate all composite gauge-invariant 
operators in terms of  Lorentz harmonic chiral superfields. 
Besides the familiar  physical fields of the $\cN = 4$ vector multiplet,  
we introduce  infinite sets of auxiliary and pure gauge fields which live in two 
gauge super-connections (harmonic superfields).
These unphysical degrees of freedom enable us to realize the chiral half of $\cN = 4$ supersymmetry off shell 
and to employ a chiral supergraph technique for perturbative calculations.
The operators, which are polynomial in terms of the usual physical fields, become nonpolynomial in terms of the harmonic superfields. 
They are represented by infinite sums of vertices of arbitrarily high valence.

Secondly, we use the harmonic super-propagators for the calculation of   MHV tree-level form factors. In this case the interaction is transferred from the Lagrangian to  the infinite number of operator vertices.
The form factors are obtained by the LSZ reduction procedure. It amounts to stretching amputated super-propagators between the on-shell states and the superfields at the operator vertices. The simplest form factors are for operators made from chiral supercurvatures only. We use the $\bQ-$transformations to reconstruct the full nonchiral multiplets of operators. The other fields from the vector multiplet are obtained by acting with spinor derivatives. After the insertion of the on-shell states in the operator the fields with dotted spinor indices are equivalently represented by space-time derivatives. This reproduces the effective operator vertex prescription of \cite{Koster:2016loo}.
Since we understand the complete construction of composite operators (unrelated to the on-shell states),  we  see that 
the effective vertices can only  work at the MHV tree level. 
At the $\rm N^k MHV$ level our supergraphs are equally applicable. For example, at the NMHV level
we have to include a vertex from the Lagrangian and to link it by a super-propagator to a  superfield from the operator. The chiral truncation of the operators 
 reproduces the CSW rules for form factors \cite{Brandhuber:2011tv}. 
 Our supergraph technique works equally well for the nonchiral operators.


\section*{Acknowledgements}

We are grateful to Grisha Korchemsky for correspondence. 
We acknowledge partial support by the French National Agency for Research (ANR) under contract StrongInt (BLANC-SIMI-4-2011). The work of D.C. has been supported by the ``Investissements d'avenir, Labex ENIGMASS'' and partially supported by the RFBR grant 14-01-00341.


\section*{Appendix} 

\appendix

\section{$\bQ-$variations of the propagator}

In this Appendix we  derive the nonchiral completion of the amputated chiral propagator
$\lim_{p^2\to0} p^2 \vev{W^{\rm free}_{++} A^{++}}$ \p{2.9'}. 
We calculate the $\bQ-$variations of the super-curvature $W_{++}$ and of 
the gauge connection $A^{++}$. This results in a finite $\bQ-$transformation which restores the $\bq$-dependence at both ends of the propagator.
In this way we derive \p{6.14}.

In view of \p{multbQ} we can throw up to four $\bQ$-variations on the propagator $\vev{W^{\rm free}_{++} A^{++}}$ and distribute them arbitrarily between its two ends. Further we apply $\bQ$-variations to the $A^{++}$ end of the propagator. 
The calculation is similar to the chiral case \p{2.5'}.
Acting with $k=1,\ldots,4$ $\bQ$-variations \p{multbQ}, which are taken in the momentum representation, and using the
propagators \p{prop1}, \p{prop2} we obtain 
\begin{align}\label{A.1}
&\vev{ W^{\rm free}_{++}(p,\q_+,v,w) \bQ_{\da_1}^{A_1} \ldots \bQ_{\da_k}^{A_k} A^{++}(-p,\q_0,u)}\\
& = 4(-\ti)^k \left[ k\, p^-_{(\da_1} \ldots p^-_{\da_{k-1}} \xi^{\-}_{\da_k)} / p^{\- +} 
+ \pi p^-_{\da_1} \ldots p^-_{\da_k} \delta^2(p^{\- +})\right]
\q^{A_1 +}_0 \ldots \q^{A_k +}_0 \delta^2(w_{+}(\q-\q_0)^+) \notag
\end{align}
where LH $v$ has dropped out and all LH projections are with $u$, i.e. $\q^{+ A} = u^+_{\a} \q^{\a A}$, $p^-_{\da} = u^{-\a} p_{\a\da}$, etc.
We are going to amputate the propagator \p{A.1}, so we need to reveal a pole $1/p^2$ in \p{A.1}.
The pole is due to the square bracket term in \p{A.1} which is a harmonic distribution. 
Like in the chiral case \p{2.8}, the pole emerges upon harmonic integration over $u$ of the distribution
with a test function.

Firstly we show that the residue of the pole is independent of the auxiliary gauge-fixing spinor $\xi^{\-}$.
This spinor is a harmonic on the factor $SU(2)_R$ of the Euclidean Lorentz group (see Sect. \ref{LHC}).
Since \p{A.1} carries zero $SU(2)_R$ harmonic charge, then  by lemma \p{5} to prove $\xi^{\-}$-independence of the residue 
we need to show that it is annihilated by $\pa^{\-\-}$.
We do the substitution $p^{-}_{\da} \to - \xi^{\+}_{\da} p^{\- -}$ for $p^{-}$'s accompanying $\delta^2(p^{\- +})$
and act on the square brackets in \p{A.1} with $\pa^{\-\-}$,
\begin{align}\label{A.2}
\pa^{\-\-}
\left[ k\, p^-_{(\da_1} \ldots p^-_{\da_{k-1}} \xi^{\-}_{\da_k)} / p^{\- +} 
+ \pi \xi^\+_{\da_1} \ldots \xi^\+_{\da_k} (-p^{\- -})^k \delta^2(p^{\- +})\right]. 
\end{align}
We simplify \p{A.2} by means of the following formula \cite{Chicherin:2016fac} 
\begin{align}\label{A.3}
\pa^{\-\-} \frac{1}{p^{\- +}} = \pi p^{\- -}\delta^2(p^{\- +})
\end{align}
so \p{A.2} is equal to 
\begin{align}\label{A.4}
\pi \xi^\+_{\da_1} \ldots \xi^\+_{\da_k} \pa^{\-\-} [(-p^{\- -})^k \delta^2(p^{\- +})]\,.
\end{align}
Then we need to show that the residue of the distribution \p{A.4} at the pole $1/p^2$ vanishes.
So we integrate \p{A.2} with a test function $\varphi^{(+k)}$, which is a holomorphic rational function of LH $u^+$ of degree $(+k)$,
by means of \p{2.8}
\begin{align}\label{}
\pi \pa^{\-\-}\int d u \, (-p^{\- -})^k \delta^2(p^{\- +}) \varphi^{(+k)}(u^+) = \frac{1}{p^2} \pa^{\-\-}\varphi^{(+k)}(p^{\-})\,.
\end{align}
On shell $p_{\a\da} = \lambda_{\a}\tilde\lambda_{\da}$, then 
$\varphi^{(+k)}(p^{\-}) = [\xi^{\-} \tilde\lambda]^k \varphi^{(+k)}(\lambda)$ is a polynomial in $\xi^{\-}$ 
and it is annihilated by $\pa^{\-\-}$. Thus we proved that the residue of \p{A.1} does not depend on the gauge-fixing $\xi^{\-}$.

In order to see explicitly how $\xi^{\-}$ drops out in the amputated \p{A.1} we need to integrate this distribution 
with a test function $\varphi^{(+k)}(u^+)$. This gives rise to a number of $\xi$-dependent terms. 
Then applying Fiertz identity multiple times we can check the cancellation of $\xi^{\dt\pm}$ among these terms.
We prefer to take a shortcut here, so we integrate the distribution over harmonics $\xi$. 
For the first term in the square brackets in \p{A.1} we integrate by parts and take into account the formula \p{A.3},
\begin{align}\label{}
\int d\xi\, \frac{\xi^{\-}_{\da}}{p^{\- +}}= \int d\xi\, \frac{\pa^{\-\-} \xi^{\+}_{\da}}{p^{\- +}}
= \pi p^{-}_{\da} \int d\xi\, \delta^2(p^{\- +}) = \frac1{p^2} p^{-}_{\da}
\end{align}
where at the last step the delta function is integrated analogously to \p{2.8}: $\pi \int d\xi\, \delta^2(p^{\- +}) = 1/p^2$.
This formula enables us to integrate the second term in the square brackets in \p{A.1}. So the sum of two terms is equal to
the averaged over $\xi$ propagator $(k+1)\, p^{-}_{\da_1} \ldots p^{-}_{\da_k}/p^2$.
Then we amputate this propagator, choose the test function to be polynomial $\varphi^{(+k)}(u^+) = (u^+ v^+_1)\ldots (u^+ v^+_k)$, 
integrate over LH $u$, and go on shell
\begin{align}\label{}
(k+1) \int d u \, p^{-}_{\da_1} \ldots p^{-}_{\da_k} \varphi^{(+k)}(u^+) 
= (p_{(\da_1} v_1^+)\ldots (p_{\da_k)} v_k^+) \ \stackrel{p=\la\tl}{\longrightarrow} 
\ \tl_{\da_1} \ldots \tl_{\da_k} \varphi^{(+k)}(\la) \,.
\end{align}
Thus the amputated propagator \p{A.1} is the following distribution 
\begin{align}\label{A.8}
&\lim_{p^2\to0} p^2 \vev{ W^{\rm free}_{++}(p,\q_+,v,w) \bQ_{\da_1}^{A_1} \ldots \bQ_{\da_k}^{A_k} A^{++}(-p,\q_0,u)}
\nt 
&=4 (-\ti)^k \tl_{\da_1} \ldots \tl_{\da_k}
\vev{\lambda \q^{A_1}_0} \ldots \vev{\lambda\q^{A_k}_0} \delta(\lambda,u) \delta^2(w_{+}(\q-\q_0) \ket{\lambda})\,.
\end{align}
It is a nonchiral deviation from the chiral formula \p{2.9'}.
The effect of the $\bQ-$variations at the $A^{++}$ end of the amputated propagator \p{2.9'} amounts to 
a number of factors $\tl_\da \vev{\la \q^A_0}$.
We remark that it is proportional to the momentum helicity spinor $\tl_\da$. Consequently, completing the variation with the antichiral odd variable $\bq_0$, we see only the projection $[\bq_{0} \tl]$ appearing. 

Some of the $\bQ-$variations in \p{A.8} can be moved to the $W_{++}$ end.
There only the RH projection $\bQ^{a'}_- = w^{a'}_{-A} \bQ^A$ gives a non-trivial result.
So we can not throw more than two $\bQ_-$ on the $W_{++}$ end.
The result is given by \p{A.8} with the right hand side projected with RH $w^{a'}_{-A}$.

So far we considered only the self-dual part of the $\bQ$-variations, 
i.e. we disregarded the $\bQ_{\rm Z}$ term in \p{bQ}. Now let us explain why it is legitimate. 
It is expressed in terms of the nonpolynomial $A^{--}$ \p{448}: 
$(\bQ_{\rm Z})^{B}_{\db} A^+_{\da} = -2(\pa^+)^4 (\q^{-B}A^{--}) \ep_{\da\db}$.
At the $W_{++}$ end of the propagator, the $\bQ_{\rm Z}$-variation can appear only in  
$\ep^{\da\db}\bar Q^{(a'}_{-\da}\bar Q^{b')}_{-\db}W^{\rm free}_{++}(p_i,\q_{i+})$ with complete $\bQ$ \p{bQ}. 
There is no such state in the spectrum, and one can show that it does not contribute in the amputated propagator.
We can also try to take $\bQ_{\rm Z}$ into account on the $A^{++}$ end, i.e. reconstruct multiplets of operators by means of 
complete $\bQ$. Owing to the L-analytic projector $(\pa^+)^4$ each $\bQ$-variation increases the Grassmann degree of the correlator by $4$ units.
So we do not need it at the MHV tree level.


\end{document}